\documentclass[preprint,aps,prd,showpacs,nofootinbib,amsmath,amssymb,amsfonts,superscriptaddress]{revtex4-1}

\usepackage{amssymb}
\usepackage{bm}
\usepackage{latexsym}
\usepackage{amsfonts,amssymb}
\usepackage{graphicx,epsfig}
\usepackage{psfrag}
\usepackage{amsmath,amssymb}
\usepackage{mathrsfs}
\usepackage{amsthm}
\usepackage{bm}
\usepackage{lipsum}
\usepackage{array}
\usepackage{tabu}
\usepackage{latexsym}
\usepackage{mathrsfs}
\usepackage{hyperref}
\usepackage{float}
\usepackage[section]{placeins}
\hypersetup{
    colorlinks=true,       
    linkcolor=red,          
    citecolor=blue,        
}
\newcommand{\addlabel}[1]{%
Eq.\refstepcounter{equation}%
\addlabel{#1}%
\let\]\endequation }

\begin{document}

\title{Low-energy modified gravity signatures on the large-scale structures}
\author{Joseph P Johnson}
\email{josephpj@iitb.ac.in}
\affiliation{Department of Physics, Indian Institute of Technology Bombay, Mumbai 400076, India} 
\author{S. Shankaranarayanan} \email{shanki@phy.iitb.ac.in}
\affiliation{Department of Physics, Indian Institute of Technology Bombay, Mumbai 400076, India}

\begin{abstract}
A large number of dark energy and modified gravity models lead to the same expansion history of the Universe, hence, making it difficult to distinguish them from observations. To make the calculations transparent, we consider $f(R)$ gravity with a pressureless matter without making any assumption about the form of $f(R)$. Using the late-time expansion history realizations constructed by Shafieloo et al~\cite{2018-Shafieloo.etal-PRD}, we explicitly show for any $f(R)$ model that the Bardeen potentials $\Psi$ and $\Phi$ evolve differently. For an arbitrary $f(R)$ model that leads to late-time accelerated expansion, we explicitly show that $|\Psi + \Phi|$ and its time-derivative evolves differently than the $\Lambda$CDM model at lower redshifts. We show that the $\Psi/\Phi$ has significant deviation from unity for larger wave-numbers. We discuss the implications of the results for the cosmological observations. 
\end{abstract}

\maketitle
\section{Introduction}

Our current understanding of the cosmos is based on an enormous extrapolation of our limited knowledge of gravity since General Relativity (GR) has not been independently tested on galactic and cosmological scales~\cite{Jain:2010ka,Will:2014kxa,Ishak:2018his}. On the largest scales, the biggest surprise from observational cosmology has been that the current Universe is accelerating~\cite{Perlmutter:1998np,Riess:1998cb}. The observations of Type Ia supernovae suggests that the current Universe is undergoing a phase of accelerated expansion \cite{2017-Scolnic.etal-ApJ} which agrees with the observations of cosmic microwave background radiation~\cite{Spergel:2006hy,Akrami:2018vks,Aghanim:2018eyx}. 

Providing a fundamental understanding of the late-time accelerated expansion of the Universe is one of the most challenging problems in cosmology. GR alone can not explain the late-time acceleration of the Universe with ordinary matter or radiation. The presence of an exotic matter source energy referred to as dark energy can explain the late-time accelerated expansion~\cite{2000-Sahni.Starobinsky-IJMPD,Copeland:2006wr,yoo2012theoretical,Tsujikawa:2013fba,2016-Davis.Parkinson-Book}. The most straightforward candidate for the dark energy (DE) is the cosmological constant  $\Lambda$~\cite{1968-Sakharov-SPD,1989-Weinberg-RMP,2003-Padmanabhan-PRep,2003-Peebles.Ratra-RMP}. However, the estimated value of $\Lambda$ from observations shows that it is many orders smaller than the vacuum energy density predicted by particle physics~\cite{1968-Sakharov-SPD,1989-Weinberg-RMP}. The cosmological constant does not change with the evolution of the Universe. However, in models like  Quintessence, K-essence, Phantom models, Chameleon scalar fields dark energy changes with time~\cite{Copeland:2006wr}.

An alternative to dark energy is the modified gravity (MG) model, where the late-time acceleration is due to the large-scale modifications to GR. Several modified gravity models, like $f(R)$, Braneworld and Galileon models, have been proposed as the possible explanation for the late-time accelerated expansion of the universe~\cite{lrr-2010-3,Nojiri:2006ri,2015-Joyce.etal-PRep,2016-Joyce.etal-ARNPC}.  Among the modified gravity models, $f(R)$ models (where $f$ is an arbitrary function of the Ricci scalar $R$) are popular owing to the simplicity of the dynamical equations. Also,  $f(R)$ models do not suffer from Ostr\"ogradsky instability \cite{2007-Woodard-Proc}.

Naturally, many phenomenological $f(R)$ models that are consistent with local gravity tests and have stable late-time de Sitter point have been proposed~\cite{2007-Hu.Sawicki-PRD,2007-Starobinsky-JETPLett,2007-Amendola.etal-PRD,2007-Appleby.Battye-PLB,2008-Tsujikawa-PRD}. These models also suffer from fine-tuning problem as like the cosmological constant. In other words, one needs to tune the threshold value of the Ricci scalar $R_0$ to obtain the observed late-time acceleration. 

As mentioned above, many different $f(R)$ models with fine-tuning can account for the late-time acceleration. Similarly, many different dark energy models, within GR, can also account for the late-time acceleration. This leads to the question: Are there signatures that distinguish dark energy and modified gravity models? 

Such parameters have been constructed in the literature~\cite{2009-Song.Koyama-JCAP}. They showed that while the background equations are degenerate, the first-order perturbations are not. In particular, Song and Koyama provided a consistency test, based on the first-order scalar metric perturbations. They proposed that the modified gravity models can be mapped to modifications in Newton's constant. They obtained two parameters that can distinguish MG and dark energy models. 

In this work, we focus on the generic $f(R)$ model, which leads to late-time expansion history that is consistent with observations. In particular, we use 6400 late-time expansion history realizations constructed by Shafieloo et al~\cite{2018-Shafieloo.etal-PRD} from the latest Pantheon supernovae distance modulus compilation~\cite{2017-Scolnic.etal-ApJ}. For each of these 6400 realizations, we obtain the evolution of $f(R)$ as a function of redshift ($z$). Using the constructed $f(R)$ in the first-order perturbation equations, we obtain observationally relevant quantities like $\Phi + \Psi$, $\Phi' + \Psi'$ and $\Psi/\Phi$.  We explicitly show that these quantities evolve differently for $f(R)$ and dark-energy models. More specifically, we show that one of the Bardeen potential $\Psi$ is suppressed compared to $\Phi$ for any $f(R)$ model that leads to late-time acceleration. To our knowledge, such an analysis has not been done earlier for an arbitrary $f(R)$. We then discuss the implication of our results relating to future observations. 

In section (\ref{sec:twoscenario}), the two scenarios --- GR  with  cosmological constant $\Lambda$ and $f(R)$ model where $f(R)$ is an arbitrary function of $R$ --- are introduced. In section (\ref{sec:Background}), we obtain the evolution of various background quantities using the model-independent data of late-time expansion history of the Universe constructed by Shafieloo et al. in \cite{2018-Shafieloo.etal-PRD,2017-LHuillier.Shafieloo-JCAP}. In section (\ref{sec:FOpert}), we obtain density perturbations and scalar metric perturbations.  In section (\ref{sec:CosmoObs}), we discuss the difference in the growth of the first-order quantities in these two scenarios and obtain the relevant variables for the cosmological observations. 
In section (\ref{sec:conclusion}), we conclude by briefly discussing the results. 

In this work we use the natural units where $c=\hbar =1$, $\kappa^2 = 8 \pi G$, and the metric signature $(-,+,+,+)$. Greek alphabets denote the 4-dimensional space-time coordinates, and Latin alphabets denote the 3-dimensional spatial coordinates. Overbarred quantities (like $\overline{\rho}(t),  \overline{f}(R), \overline{F}(R)$) are evaluated for  the FRW background. $H_0$ is the Hubble constant and it does not explicitly appear in the final evolution equations for both the scenarios. For matter density parameter $\Omega_m$, we use the value calculated from the PLANCK data~\cite{Akrami:2018vks,Aghanim:2018eyx}. Unless otherwise specified, $\textit{prime}$ denotes the derivative w.r.t. to redshift $z$.

\section{Framework and the two scenarios}\label{sec:twoscenario}

As mentioned in the introduction, we consider two scenarios --- Dark Energy and $f(R)$ gravity --- that explain the late-time acceleration of the Universe. 
In this section, we briefly discuss the two scenarios and use the expansion history realizations constructed by Shafieloo et al~\cite{2018-Shafieloo.etal-PRD,2017-LHuillier.Shafieloo-JCAP} to obtain the evolution of $f(R)$ as a function of $z$.
\begin{enumerate}
\item {\bf Scenario I:}  In this scenario, we consider General Relativity with dark energy,  where dark energy is represented by cosmological constant $\Lambda$. The field equations are given by:
\begin{equation}
\label{eq:GRfieldeq}
R_{\mu \nu}-\dfrac{1}{2}g_{\mu \nu}R + \Lambda g_{\mu \nu}=\kappa^2 T_{\mu \nu} \, ,
\end{equation}
where $T_{\mu \nu}$ is the stress-tensor of the matter fields. We consider only the pressureless matter while keeping the successes of standard cosmology at early times.

\item {\bf Scenario II:}  In this scenario, we assume that the late-time acceleration 
is due to the large-scale modifications to GR which is given by $f(R)$ where $f$ is a continuous, and arbitrary function of $R$. The action and the corresponding field equations are given by: 
\begin{eqnarray}
\label{eq:fRaction}
& & S_{II} = \dfrac{1}{2 \kappa^2} \int d^{4} x \sqrt{-g} \, f(R) + \int d^{4} x \sqrt{-g} \,  \mathcal{L_{M}} \\
\label{eq:fRfieldeq}
& & F R_{\mu \nu} - \dfrac{1}{2} f(R) g_{\mu \nu} - \nabla_{\mu} \nabla_{\nu} F + g_{\mu \nu} \Box F = \kappa^2 T_{\mu \nu},
\end{eqnarray}
where $F=\dfrac{\partial f}{\partial R}$. In the case of $f(R)$ gravity, unlike General Relativity, the trace of the field equation (\ref{eq:fRfieldeq}) is dynamical~\cite{2007-Woodard-Proc}:
\begin{equation}
\label{eq:Trace}
R \, F(R) +3 \, \square F(R) -2 \, f(R) = \kappa^2 T
\end{equation}
Thus, in $f(R)$ gravity, the scalar curvature $R$, which can be expressed in terms of the metric and its derivatives, plays a non-trivial role in the determination of the metric itself.  As a result, $f(R)$ gravity has 11 dynamical variables --- 10 metric variables ($g_{\mu\nu}$) and $F(R)$. Note that the above equation points that $F(R)$ is a dynamical quantity as $F(R)$ is acted on by the differential operators. 

Here again, we consider only the pressureless matter, while keeping the success of standard general relativity at early times. In other words, we assume that until around the redshift of $1.2$ the Universe can be described by GR with a dominant contribution from the pressureless matter.
\end{enumerate}

To distinguish the above two scenarios in a model-independent manner, we do not assume any form of $f(R)$, i. e., $f(R)$ is an arbitrary function.  Instead, we assume that both scenarios lead to the same background evolution of the Universe and have the same evolution of the Hubble parameter ($H(z)$). In other words, the input parameter for the two scenarios is $H(z)$ as a function of redshift $z$. We use the model-independent data constructed by Shafieloo et al. in \cite{2018-Shafieloo.etal-PRD,2017-LHuillier.Shafieloo-JCAP}. In particular, we use 6400 late-time expansion history realizations constructed by Shafieloo et al~\cite{2018-Shafieloo.etal-PRD} from the latest Pantheon supernovae distance modulus compilation~\cite{2017-Scolnic.etal-ApJ}. 

It is important to note that the Pantheon dataset consists of $1048$ supernovae in the redshift range $[0.01,  2.26]$. Pantheon dataset has $630$, $832$, and $1025$ supernovae below $z = 0.3$,  $z = 0.5$ and $z = 1$, respectively~\cite{2017-Scolnic.etal-ApJ}. The data is sparse beyond the redshift of $1.2$, leading to uncertainty in the determination of $H(z)$ beyond the redshift of $1.2$~\cite{2018-Shafieloo.etal-PRD}. Note that strong deviations from $\Lambda$CDM are allowed by the data at $z \gtrsim 1$ in the reconstructed Hubble parameter ~\cite{2019-LHuillier.etal-Mon.Not.Roy.Astron.Soc.}. Hence, in our analysis, for the 6400 late-time expansion history realizations, we consider the evolution of the Hubble parameter in the range $0.01 < z < 1.2$. 

In the next section, using the above realizations, we obtain $\overline{F}$ as a function of $z$. In Sec. (\ref{sec:FOpert}), we use the evolution of $\overline{F}(z)$ to obtain the first-order scalar perturbations in both the scenarios. {We use the evolution of $H(z)$ and $\overline{F}(z)$ to obtain first-order scalar perturbations in both the scenarios. }

\section{Background evolution in the two scenarios}
\label{sec:Background}

In this work, we consider spatially flat FRW line-element: 
\begin{equation}
ds^2 = -dt^2 + a^2(t) \, \delta_{ij} \, dx^i dx^j \, , 
\end{equation}
where $a(t)$ is the scale factor, $\delta_{ij}$ is the Kronecker delta. As mentioned in Sec. (\ref{sec:twoscenario}), for both the scenarios, the matter content of the Universe is pressureless dust.

For the above background, the time component of the divergence of both Eqs.~(\ref{eq:GRfieldeq}) and (\ref{eq:fRfieldeq}) results in the continuity equation satisfied by the matter energy density~\cite{1993-Hamity.Barraco-GRG,2016-Tian-GRG}. For the pressureless dust, written in terms of redshift ($z$), the conversation equation leads to:
\begin{equation}
\label{eq:rhocon}
(1+z)\frac{d \overline{\rho}(z)}{dz} - \overline{\rho}(z) = 0 .
\end{equation}
Thus, the evolution of background energy density $\overline{\rho}(z)$ is identical in both the scenarios and does not explicitly depend on $h(z)$. The evolution of $\overline{\rho}(z)$ is given by
\begin{equation}
\label{eq:rhosol}
\overline{\rho}(z) = \rho_0 (1+z)^3, \quad \rho_0 = \dfrac{3  H_0^2  \Omega_m}{\kappa^2}.
\end{equation}

It is possible to obtain the evolution of background matter energy density $\bar{\rho}(z)$ using a different procedure. For instance, in the case of Scenario I, the field equations (\ref{eq:GRfieldeq}) lead to:
\begin{equation}
H^{2}(z) = \frac{\Lambda}{3} + \frac{\kappa^2}{3} \overline{\rho}(z)
\label{eq:Friedmann}
\end{equation}
where $\bar{\rho}(z)$ is the background matter density. Using he value of cosmological constant to be $\Lambda =  (4.24 \pm 0.11) \times
10^{-66} eV^2$ from the PLANCK-2018~\cite{Aghanim:2018eyx} and the value of $h(z) \equiv H(z)/H_0$ from the late-time expansion history by Shafieloo et al~\cite{2018-Shafieloo.etal-PRD}. Substituting these two quantities in the above Friedmann equation, we obtain the evolution of the background matter energy density $\bar{\rho}(z)$.

\subsection{Scenario I}

For this scenario, exact analytical solution for the scale factor exists~\cite{2000-Sahni.Starobinsky-IJMPD}:
\begin{equation}
\label{eq:LCDMExact}
H(t) = \frac{2  \alpha }{3} \, \coth(\alpha t) \, ; \quad  \alpha = \sqrt{\frac{3 \Lambda}{4}} \, ; 
\quad \bar{\rho}(t) = \frac{\rho_0}{a^3(t)} = \rho_0 (1 + z)^3 \, ; 
\quad \rho_0 = \frac{\Lambda}{2 \kappa^2}
\end{equation}

However, in this work, we will use the evolution of the matter density to be given by Eq.~(\ref{eq:rhosol}). We will use the value of matter density parameter $\Omega_m = 0.3158$, obtained from the Planck-2018 analysis \cite{Aghanim:2018eyx}.

\subsection{Scenario II} 
For the FRW background, the field equations (\ref{eq:fRfieldeq}) 
lead to 
\begin{equation}
\label{eq:FRequation}
H^{2}(z) \frac{d^{2} F(z)}{d z^{2}}+\left(H(z)  \frac{d H}{d z} +2 \frac{H^{2}(z)}{(1+z)} \right) \frac{d F(z) }{d z} 
-2 \frac{H(z)}{(1+z)}  \frac{d H}{d z} F(z) + 
\frac{\kappa^{2}}{(1 + z)^2} \bar{\rho}(z) =0
\end{equation}
As mentioned above, due to the sparsity of the PANTHEON data beyond the redshift of $1.2$~\cite{2017-Scolnic.etal-ApJ}, we assume that beyond the redshift of $1.2$ the Universe can be described by {GR with cosmological constant} with dominant contribution from the pressureless matter $\bar{\rho}(z)$. In other words, the modifications to gravity {begin to} dominate the evolution of the Universe around $z \sim 1.2$. In order to obtain the evolution of $F(z)$ as a function of $z$ in the above equation, we use the following initial conditions:
\begin{equation}
\label{eq:InitialCondition}
{F}(z=1.2)=1, \quad  \mbox{and} \quad 
{\frac{dF}{dz}} \Big|_{z=1.2} =10^{-5}
\end{equation}
We would like to mention the following points regarding the initial conditions: First, the analysis is independent of the choice of the initial value of $z$. Our choice of the 
initial value of $z$ is linked to the data set. Second, the condition ${F}(1.2) = 1$ implies that at $z = 1.2$, the gravity is described by General Relativity. The condition  ${dF}(z = 1.2)/dz = 10^{-5}$ provides the initial choice of the rate of change of $F$. For the above initial conditions, 6400 late-time expansion history realizations constructed by Shafieloo et al~\cite{2018-Shafieloo.etal-PRD} gives the evolution of $h(z)$ and, evolution of $\overline{\rho}$ is given by Eq.(\ref{eq:rhosol}). Using these quantities , we obtain $F(z)$ as a function of $z$ using the mid-point discretization (\ref{eq:centaldiff2}) in Eq. (\ref{eq:FRequation}).
\begin{figure}[!htb]
\begin{minipage}[b]{.5\textwidth}
\includegraphics[scale=0.25]{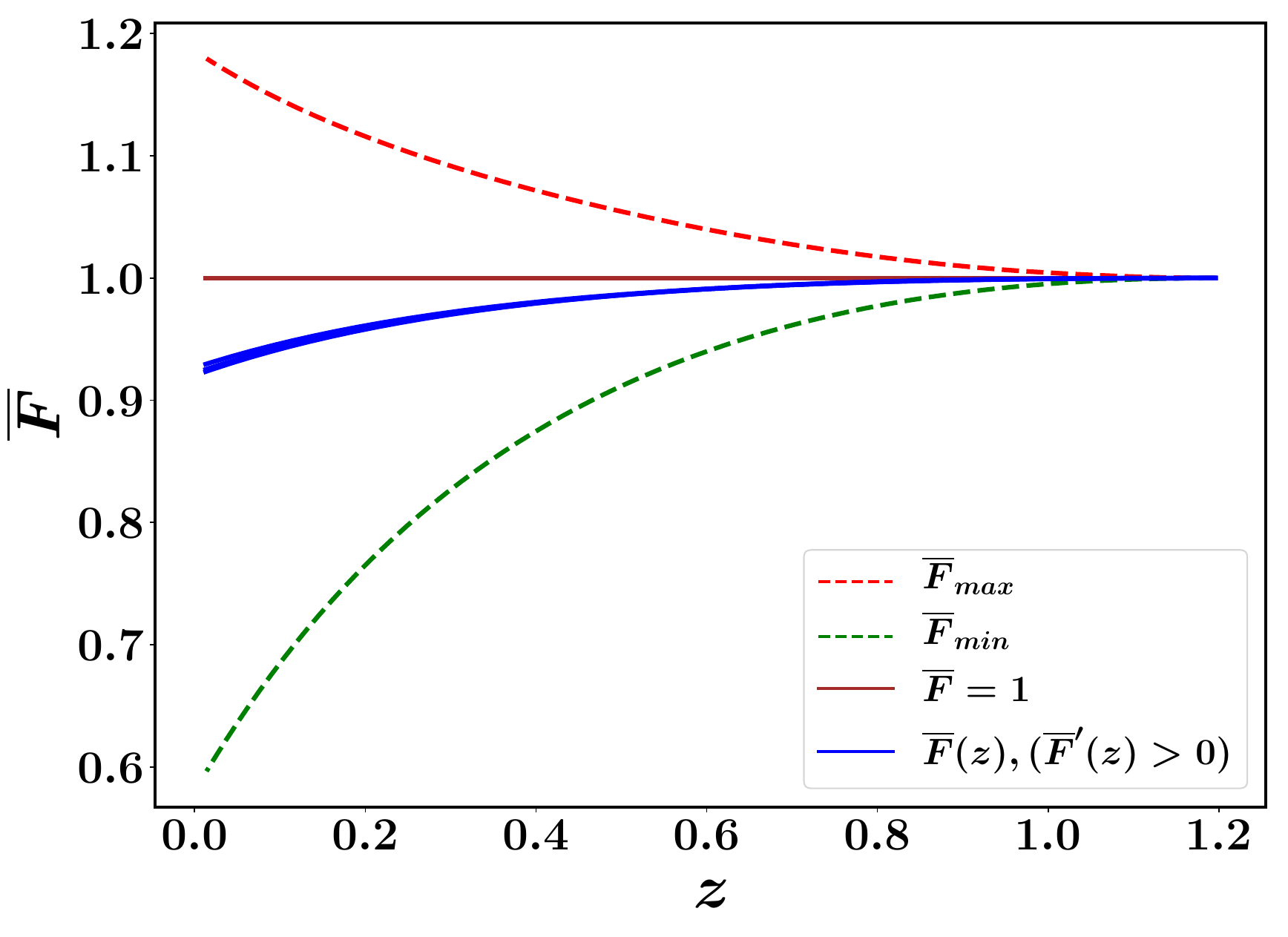}
\end{minipage}\hfill
\begin{minipage}[b]{.5\textwidth}
\includegraphics[scale=0.25]{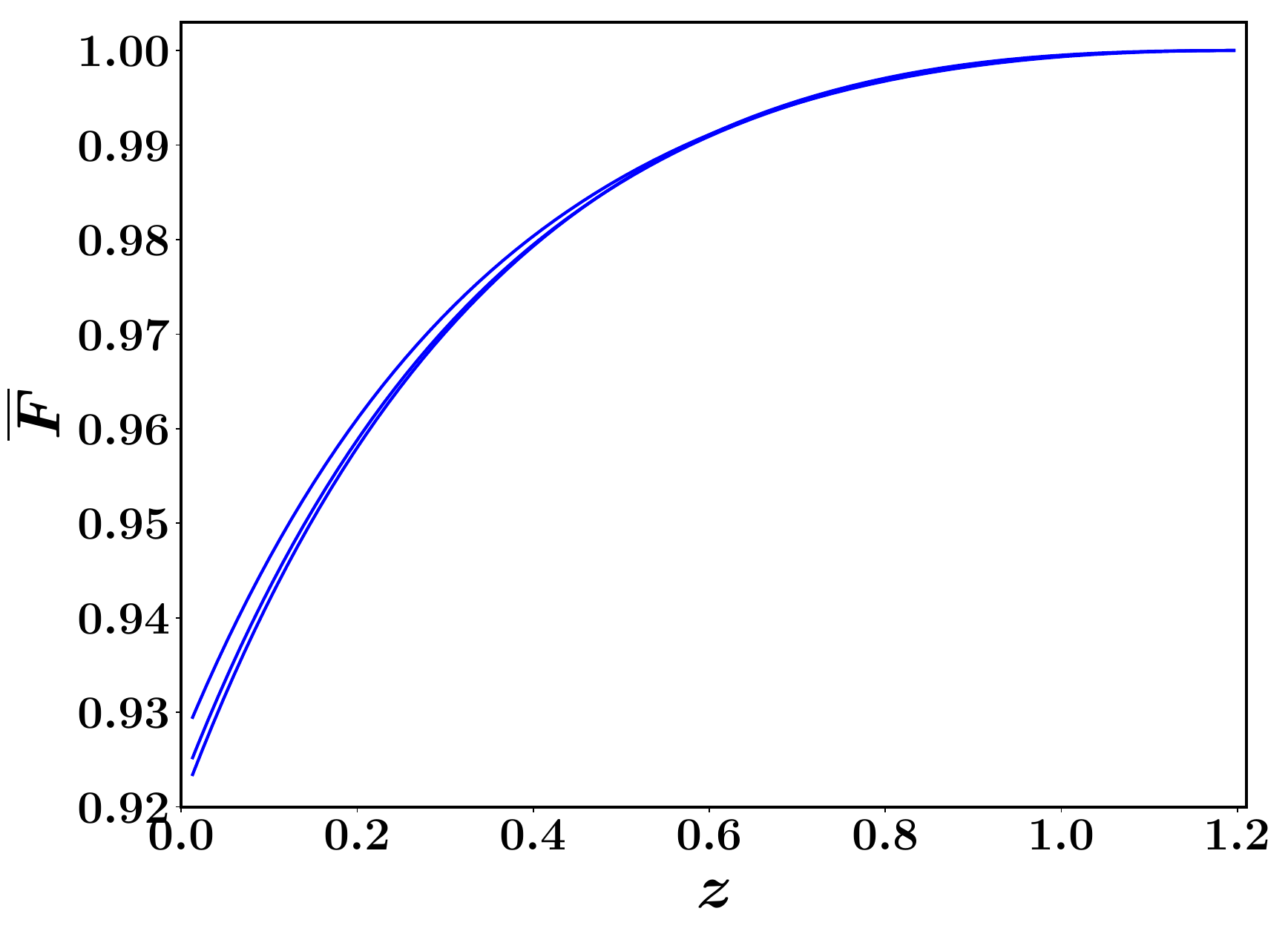}
\end{minipage}
\caption{Evolution of $F$ as function of redshift $z$: For all the datasets (Left panel) and for datasets satisfying the condition $F'(z)>0$ for all $z$(Right panel).}
\label{fig:F_evo}
\end{figure}
Evolution of $F(z)$ corresponding to the 6400 realizations of the expansion history of the universe lies with in the dashed lines in the left panel of Fig.(\ref{fig:F_evo}). Blue lines in both panels represent the evolution of $F$ corresponding to the datasets which satisfy the condition $F'(z)>0$ for all values of $z$. This condition is imposed to avoid any singularities in the evolution equations of scalar metric perturbations. For $\Omega_m = 0.3158$, there are 4 such datasets. In the following sections, we consider only these datasets for the analysis. {The number of data-sets that satisfy this condition depends on the value of $\Omega_m$ and the initial value of $dF/dz$}

This is the first key result regarding which we would like to stress the following points: 
First,  various realizations of the expansion history of the Universe corresponds to a wide range of evolution of the function $F(z)$. 
But all except a few of those solutions will result in singularities in the evolution of the scalar perturbations. Second, the above analysis does not assume any form of $f(R)$. It is possible that many different $f(R)$ models, with fine-tuned parameters, may produce the same evolution. In Appendix (\ref{sec:fit}), we have used the popular $f(R)$ models~\cite{2007-Hu.Sawicki-PRD,2007-Starobinsky-JETPLett,2007-Amendola.etal-PRD,2007-Appleby.Battye-PLB,2008-Tsujikawa-PRD} that lead to the late-time accelerated expansion and how they compare with the generic $F(z)$. Third, the evolution of $F(z)$ does not depend on the value of $dF/dz$ at $z = 1.2$. Appendix (\ref{sec:Fpini}) contains the plots of evolution of $F(z)$ for different values of $dF/dz$ at $z = 1.2$. These plots clearly show that the evolution of $F(z)$ is independent of the initial condition on $dF/dz$.

\section{First order scalar perturbations in the two scenarios}
\label{sec:FOpert}
We aim to distinguish between GR and modified gravity models using observations. To obtain the physical parameters that can be used to separate the two scenarios, we need to obtain perturbed quantities about the FRW background.  

In the largest scales, it is a good approximation to assume that the perturbed part is  small compared to the background. More specifically, the perturbed energy density is smaller than the (average) background density. The {first order} scalar perturbations about the FRW line-element in the Newtonian gauge is given by~\cite{2005-Mukhanov-Book}: 
\begin{equation}
\label{eq:FRWpert}
 ds^2=-(1+2 \Phi)dt^2 + a^2(t) \, (1-2 \Psi) \delta_{i j}dx^i dx^j \, ,
\end{equation}
where $\Phi \equiv \Phi(t,x^i)$ and $\Psi \equiv \Psi(t,x^i)$ are the scalar perturbations.  As mentioned earlier, for both the scenarios, the matter 
content of the universe is represented by pressureless dust with 
energy momentum tensor:
\begin{equation}
\label{eq:emtensor1}
T_{\mu \nu}= (\rho + p) \, u_{\mu} \, u_{\nu}+p  \, g_{\mu \nu}
\end{equation}
where $u_{\mu}$ is the four velocity, 
$\rho(t, x^i) = \bar{\rho}(t) \left(1 + \delta(t, x^i) \right)$  is the energy density including the first order density perturbations, $\delta(t, x^i) =  \delta \rho(t, x^i)/\bar{\rho}(t)$ is the fractional amplitude of density perturbations, and $p$ is the pressure of the fluid which is taken to be zero.
\subsection{Scenario I}

The first order perturbed Einstein's equation in this Scenario leads to the following equations in the Fourier space~\cite{2005-Mukhanov-Book}:
\begin{eqnarray}
\label{eq:deltaevodustGR}
 \ddot{\delta}_{\rm GR} +2H\dot{\delta}_{\rm GR} -\dfrac{\kappa^2}{2} \, \bar{\rho} \, \delta_{\rm GR} &=& 0 \\
 \label{eq:Phidelta}
 \frac{k^2}{a^2} \, \Phi_{\rm GR} +\dfrac{\kappa^2}{2} \, \bar{\rho} \, \delta_{\rm GR}  &=& 0 \\ 
\label{eq:phieqpsi}
 \Phi_{\rm GR} - \Psi_{\rm GR} &=&0.
\end{eqnarray}
Even though the result for this scenario is trivial, the procedure we follow is the same for both the scenarios: First, using the background density $\overline{\rho}(z)$ in Eq. (\ref{eq:deltaevodustGR}), we obtain the fractional amplitude of density perturbations $\delta_{\rm GR} (z)$. [We use the expressions in Appendix (A1), to convert the differential equations from $t$ to $z$.] Next, substituting  $\delta_{\rm GR} (z)$ in Eq. (\ref{eq:Phidelta}), we obtain $\Phi_{GR}$. In the case of GR, with single fluid, the two Bardeen potentials ($\Phi_{GR}, \Psi_{GR}$) are identical. 

\subsection{Scenario II}

As mentioned earlier, the scalar curvature $R$ satisfies the differential equation (\ref{eq:Trace}) and plays a non-trivial role in the determination of the metric itself.  
Hence, $F(R)$ can be treated as a dynamical variable.  For the perturbed FRW line-element (\ref{eq:FRWpert}), $i \neq j$ component of the modified Einstein's equations (\ref{eq:FRequation}) leads to:
\begin{equation}
\label{eq:PhIPsi-fR}
\Phi_{\rm MG} - \Psi_{\rm MG} = - \frac{\delta F}{\overline{F}} \quad 
\mbox{where} \quad \delta F = \overline{\left( \frac{\partial F}{\partial R} \right)} \delta R 
\end{equation}
The above expression shows that the two Bardeen potentials ($\Phi_{\rm MG}, \Psi_{\rm MG}$) are not identical and their difference depends on $F(R)$. [We only list the key equations in this section and relegate the details to Appendix (\ref{app:Scenario2}).] 

\begin{figure}[!h]
\includegraphics[scale=0.45]{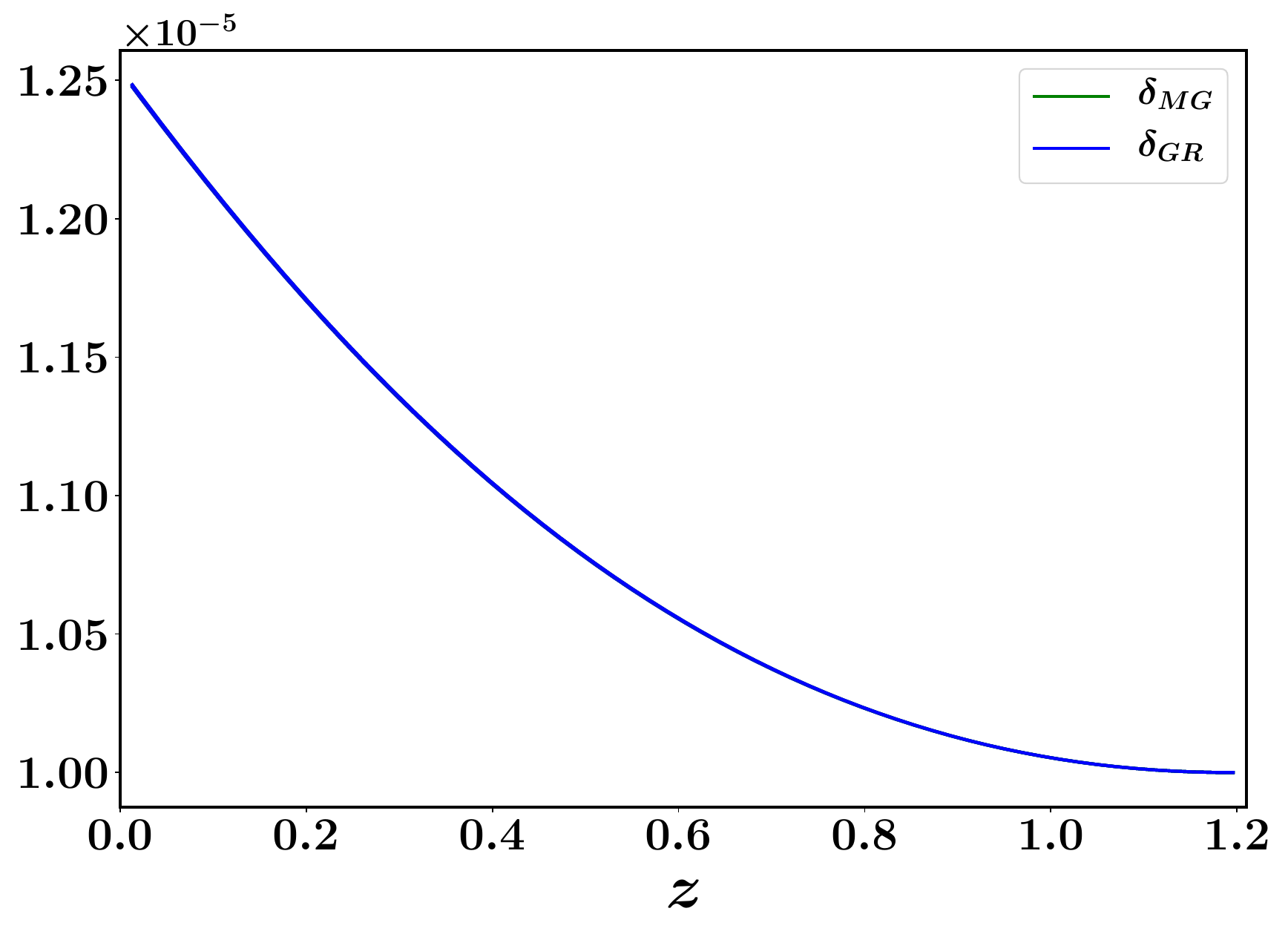} 
\caption{Evolution of $\delta_{\rm MG}$ and $\delta_{\rm GR}$ as a function of $z$ for $k = H_0$, and all realizations.}
\label{fig:delta12}
\end{figure}

The evolution of fractional amplitude of density perturbations $\delta(z)$ in $f(R)$ model is given by~\cite{2009-Gannouji.etal-JCAP,2018-Akarsu.etal-EPJC}:
\begin{equation}
 \label{eq:deltaevodustMG}
\ddot{\delta}_{\rm MG} +2H\dot{\delta}_{\rm MG} - \frac{\kappa^2_{\rm eff}}{2} \,  \bar{\rho} \, \delta_{\rm MG} \, = \, 0 
\end{equation}
where 
\begin{equation}
\kappa^2_{\rm eff}=  \dfrac{\kappa^2}{F} \left(1+4 \frac{k^{2}}{a^2} 
\frac{\partial \overline{\ln F}}{\partial R}\right) \Big/ \left(1+3 \frac{k^{2}}{a^2} 
\frac{\partial \overline{\ln F}}{\partial R} \right)
\end{equation}
Note that in the limit of $f(R) \to R$, $\kappa_{\rm eff} \to \kappa$ and 
$\delta_{\rm MG} \to \delta_{\rm GR}$. Like in the earlier scenario, using the background density $\overline{\rho}(z)$, we obtain the fractional amplitude of density perturbations $\delta_{\rm MG}(z)$. The evolution of $\delta_{GR}$ and $\delta_{MG}$ is plotted in Fig.(\ref{fig:delta12}) .

As we see here, for the data sets considered, the evolution of $\delta_{GR}$ and $\delta_{MG}$ are near identical.

\begin{figure}[!h]
  \begin{minipage}[b]{0.5 \textwidth}
\includegraphics[scale=0.25]{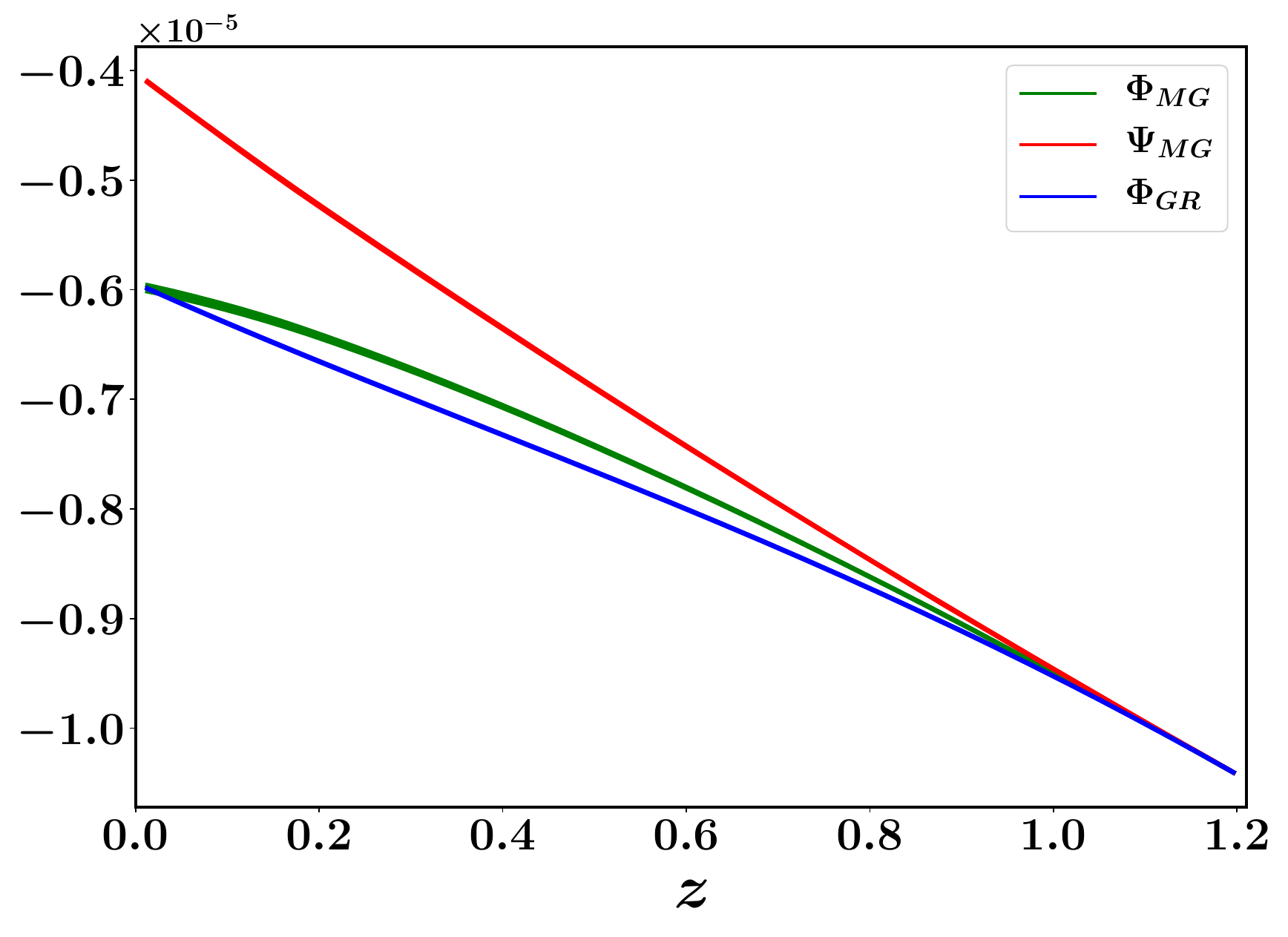} 
\end{minipage}\hfill
 \begin{minipage}[b]{0.5 \textwidth}
 \includegraphics[scale=0.25]{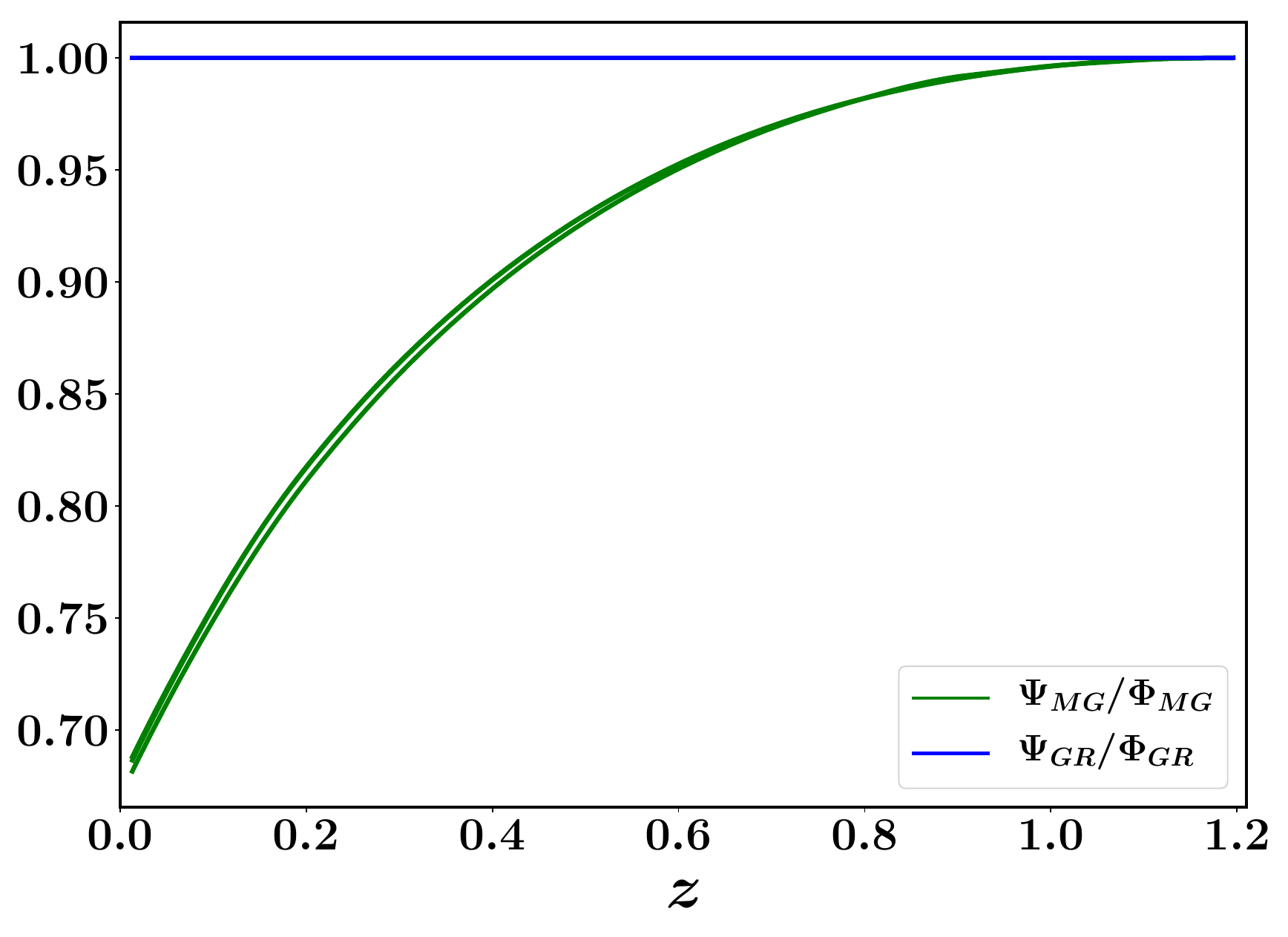}
 \end{minipage}
\caption{Evolution of $\Phi_{\rm MG}$, $\Psi_{\rm MG}$ and $\Phi_{\rm GR}$(left panel) and the evolution of $\Psi_{\rm MG}/\Phi_{\rm MG}$ and $\Psi_{\rm GR}/\Phi_{\rm GR}$(right panel) as a function of redshift $z$ for $k = H_0$.}
\label{fig:Psi_Phi}
\end{figure}

The Bardeen potentials satisfy the following coupled differential equations:
\begin{eqnarray}
\label{eq:Psimg}
\dot{\Psi}_{\rm MG} + 
\left(H-\frac{F \dot{H}}{\dot{F}}+\frac{F}{3 \dot{F}}\frac{k^2}{a^2} \right)\Phi_{\rm MG} +  \left(\frac{F \dot{H}}{\dot{F}}+\frac{F}{3 \dot{F}}\frac{k^2}{a^2}\right)\Psi_{\rm MG} 
 +\frac{\kappa^2  \overline{\rho} }{3 \dot{F}} \delta_{\rm MG} &=& 0 \\
 \label{eq:Phimg}
\!\!\!\!\!\!\!\!\!\!\!\!\!\!\!\! \dot{\Phi}_{\rm MG} + 
 \left(H-\frac{\dot{F}}{F} - \frac{F \dot{H}}{\dot{F}} - 
 \frac{F}{3 \dot{F}}\frac{k^2}{a^2} \right) \Psi_{\rm MG} +
 \left(2 \frac{\dot{F}}{F}+\frac{F \dot{H}}{\dot{F}}-\frac{F}{3 \dot{F}}\frac{k^2}{a^2}\right) \Phi_{\rm MG} 
 - \frac{\kappa^2 \overline{\rho}}{3 \dot{F}} \delta_{\rm MG} &=&0
\end{eqnarray}
Using $\delta_{\rm MG}$ from Eq. (\ref{eq:deltaevodustMG}), we numerically solve the above differential equations for the four realizations which satisfies the condition $F'(z)>0$ and is plotted in Fig.~(\ref{fig:Psi_Phi}). 

\begin{figure}[!h]
  \begin{minipage}[b]{0.5 \textwidth}
\includegraphics[scale=0.25]{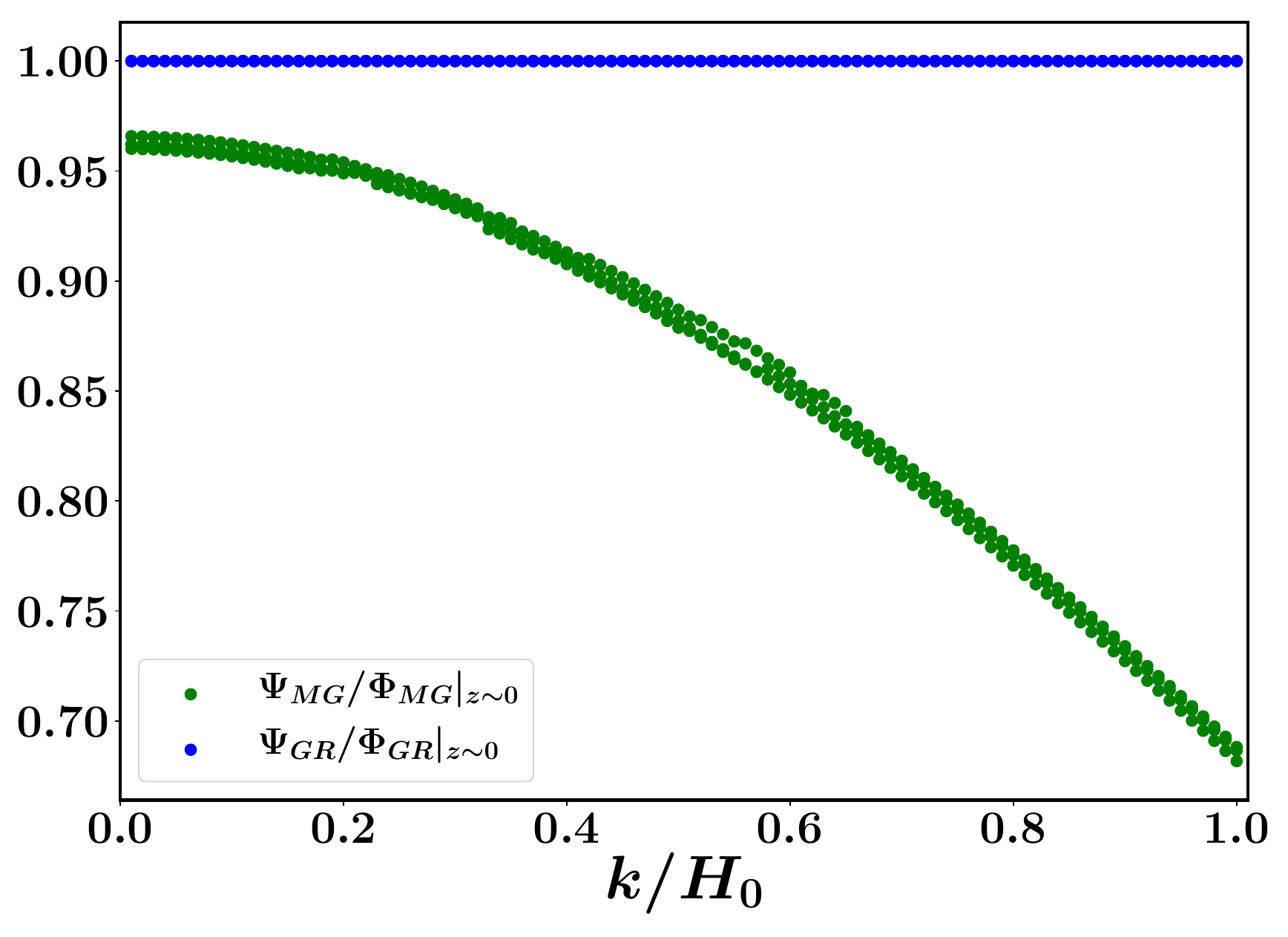} 
\end{minipage}\hfill
 \begin{minipage}[b]{0.5 \textwidth}
 \includegraphics[scale=0.25]{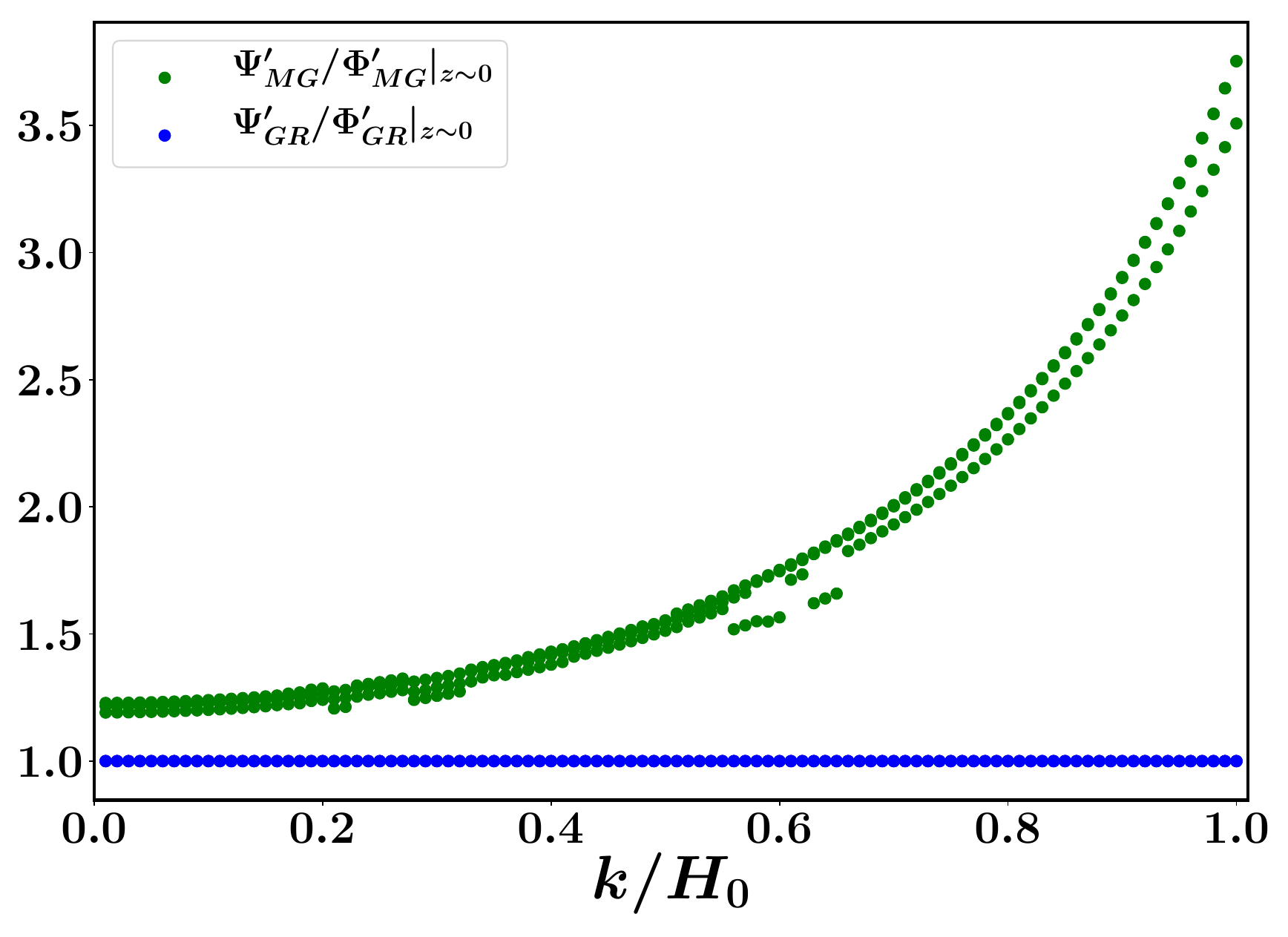}
 \end{minipage}
\caption{Value of $\Psi / \Phi$(left panel) and $\Psi^{\prime} / \Phi^{\prime} $(right panel) at $z \sim 0$ in both the scenarios as a function of $ k/H_0$}
\label{fig:psi_by_psi_last}
\end{figure}

This is the second key result regarding which we would like to stress the following points: First, for any arbitrary $f(R)$ that leads to late-time acceleration $\Phi_{\rm MG} \neq \Psi_{\rm MG}$ and the evolution of $\Psi_{\rm MG}/\Phi_{\rm MG}$ and $\Psi^{\prime} / \Phi^{\prime} $  deviates from the scenario I at the lower redshifts. Specifically, we have shown that at late-times $\Psi_{\rm MG}$ is less than $\Phi_{\rm MG}$.

Second, the deviation of $\Psi_{MG}/\Phi_{MG}$ also depends on the value of the scaled wave number $\ k/H_0$, as we see in Fig.~(\ref{fig:psi_by_psi_last}). During the course of the evolution, perturbation modes with larger wave number shows larger deviation of $\Psi_{MG}/ \Phi_{MG}$ from $\Psi_{GR} / \Phi_{GR} = 1$. This implies that the scalar perturbations with largest possible length scales $(k/H_0 \to 0)$ are not affected by $f(R)$. However, the perturbations within the current horizon radius are affected by $f(R)$.  Our results are consistent with the Universe initially underwent inflation, followed by the standard model of cosmology of the Universe.  The longer wavelength modes during inflation leave the Hubble radius at earlier epochs, and these modes reenter the current epoch much later and have undergone little structure formation. Thus, the modes within the event-horizon have been affected by the modified gravity while the longer wavelength modes are not affected.

\section{Confronting with observations}
\label{sec:CosmoObs}

In the previous section, we showed that even if both the scenarios lead to the same background evolution, the scalar perturbations in both scenarios evolve differently. 

\begin{figure}[!h]
\includegraphics[scale=0.45]{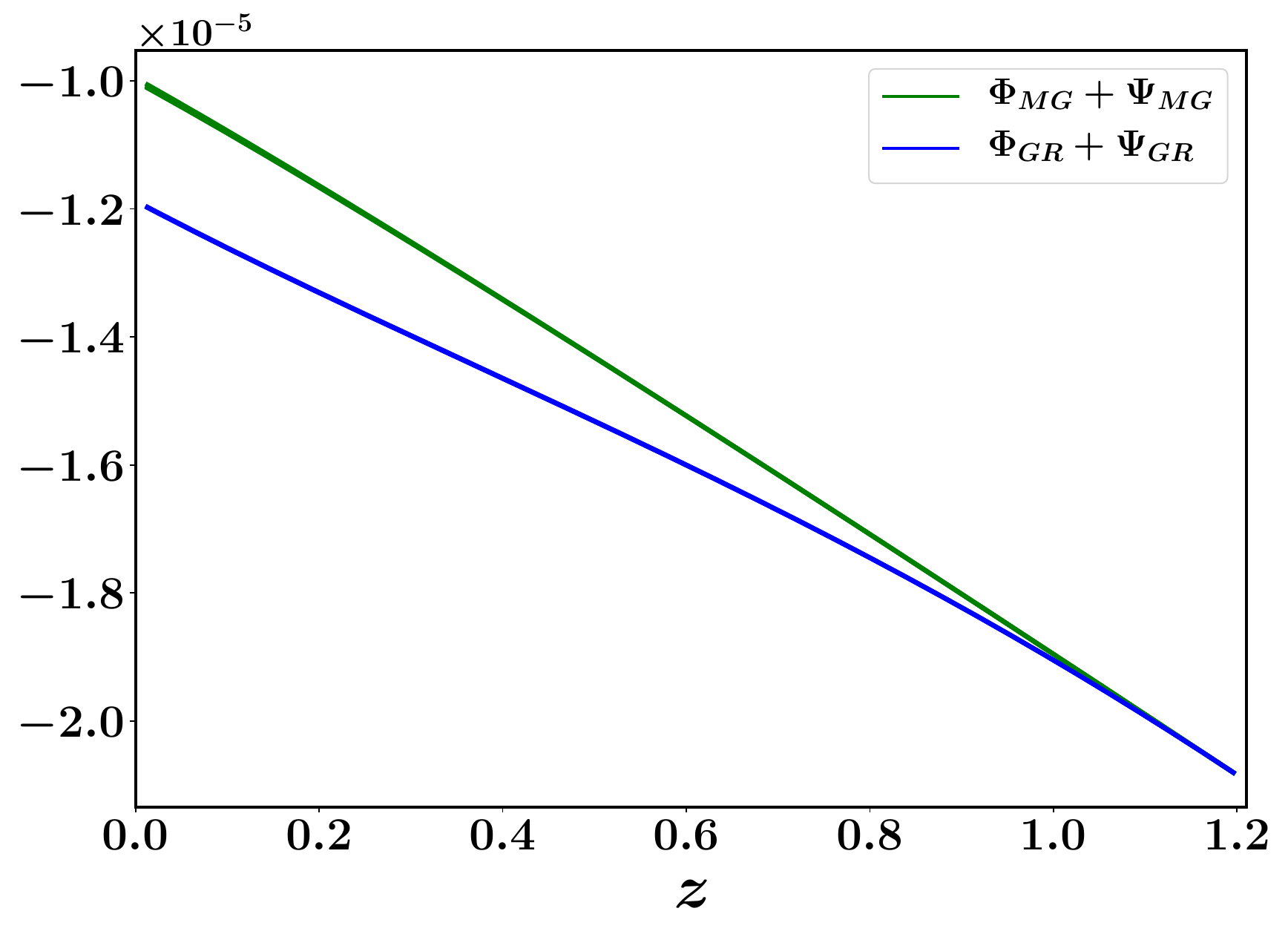}
\caption{Evolution of $\Psi_{MG}+\Phi_{MG} $ and $ \Psi_{GR}+\Phi_{GR}$ as a function of redshift $z$  for $k=H_0$. }
\label{fig:PhiplusPsi}
\end{figure}

The quantity $\Phi+\Psi$ determines the geodesic of a photon, which affects the weak gravitational lensing~\cite{2005-Mukhanov-Book}. Figure (\ref{fig:PhiplusPsi}) contains the evolution of $\Phi+\Psi$ for both the scenarios. This is the third key result regarding which we would like to stress the following: 

In the case of GR, the quantity $|\Psi +\Phi|$ is larger as compared to the $f(R)$ in the current epoch for the observed matter density, and the relative difference is of the order of $0.1$. Though the difference is small, including reliable high redshift data beyond $z > 1.1$ might provide a better estimate of the difference between GR and $f(R)$ models. 

Even though this is not much of a difference, as we mentioned earlier, the choice of redshift range over which these quantities are evolved was made based on the availability of the reliable observational data at higher redshifts.

\begin{figure*}[!h]
\centering
\includegraphics[scale=0.45]{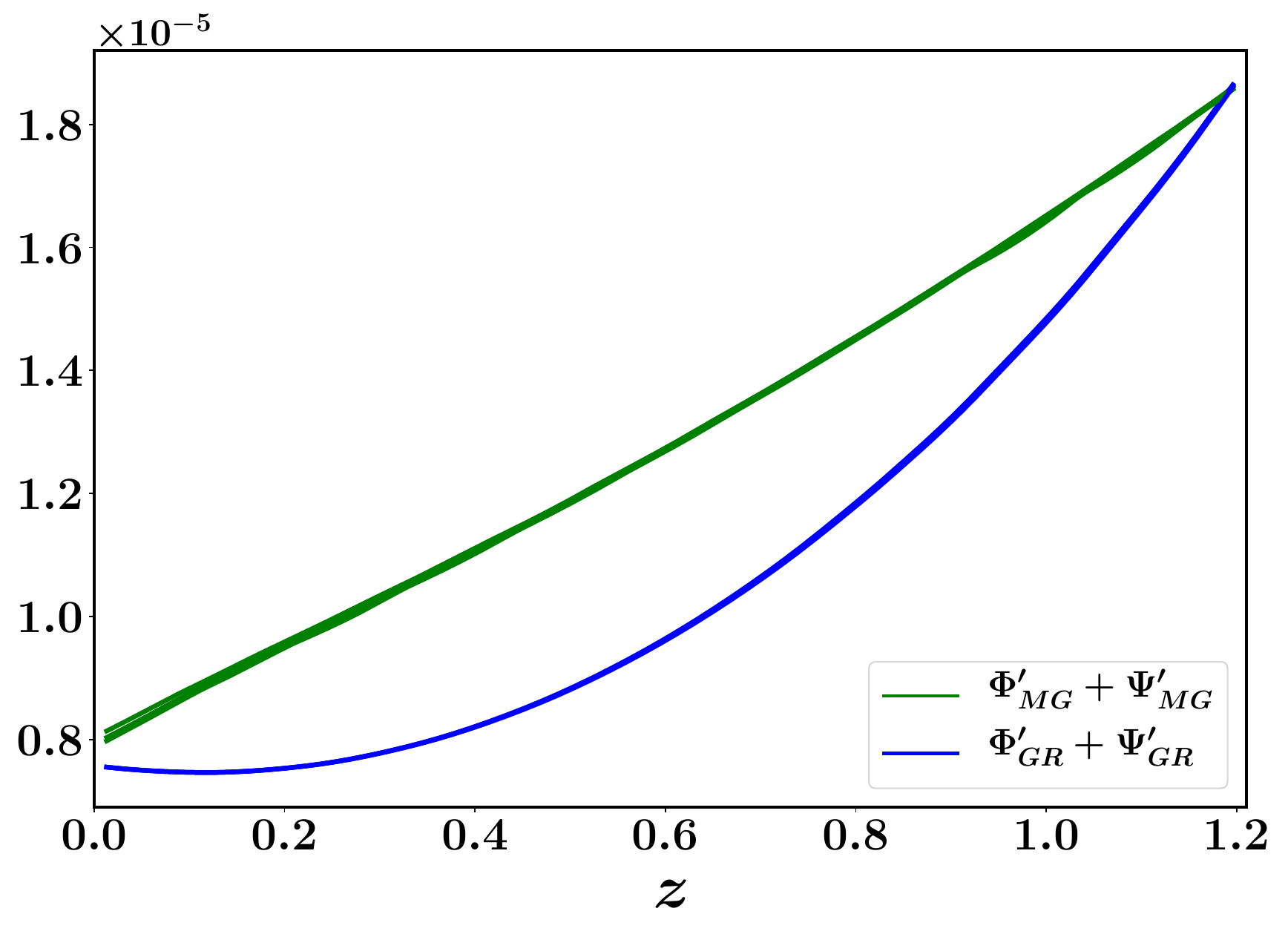}
\caption{Evolution of $\Psi_{\rm MG}^{\prime}+\Phi_{\rm MG}^{\prime} $ and  $ \Psi_{\rm GR}^{\prime}+\Phi_{\rm GR}^{\prime}$ as a function of redshift $z$  for $k=H_0$.}
\label{fig:Psi_plus_Phi_prime}
\end{figure*}

Since the Bardeen potentials evolve differently in case of $GR$ and $f(R)$, this change should potentially change the temperature fluctuations of the CMB photons. In other words, the rate of change of the $(\Phi+\Psi)$ w. r. t. $\eta$ contribute to the evolution of scalar perturbations in the temperature fluctuations in CMB in large scales --- Integrated Sachs Wolfe effect~\cite{Gorbunov:2011zzc}. Here $\eta$ is the conformal time which is related to the cosmic time via $\eta = \int dt/a$. Figure (\ref{fig:Psi_plus_Phi_prime}) shows the $\Phi^{\prime}+\Psi^{\prime}$ as a function of $z$ where \textit{prime} denotes the derivative w.r.t. $\eta$.

The two plots show the difference in the evolution of $\Phi + \Phi$ and $\Phi^{\prime} + \Psi^{\prime}$ in both the scenarios. Our analysis provides a possibility to distinguish the two scenarios using weak gravitational lensing and Integrated Sachs Wolfe effect in a model-independent manner. Lack of reliable data at higher redshifts prevents us from making a precise quantitative prediction of the differences in the evolution of scalar perturbation.  Although the relative differences in the values of these quantities at lower redshifts in both the scenarios are of the order of $10^{-1}$, evolution over a larger redshift range might lead to more significant differences in the evolution of these quantities in the two scenarios considered. To demonstrate this, we have evolved the relevant quantities from $z=1.5$ to $z \sim 0$, and the results show a more significant difference in the evolution of the $\Phi^{\prime} + \Psi^{\prime}$ in the two scenarios. See Appendix (??) for details. 

\section{Conclusions}
\label{sec:conclusion}

In this work, we have investigated in detailed the two scenarios --- General Relativity with cosmological constant and $f(R)$ gravity --- which can explain the late-time acceleration of the Universe. We have shown that in these two scenarios for which the background evolution is identical, the growth of scalar perturbations is different. More specifically, we have demonstrated that at late-times $\Psi_{\rm MG}$ is less than $\Phi_{\rm MG}$. 
We have shown that the difference in the growth of the scalar perturbations can be used to distinguish the two scenarios using the weak lensing and Integrated Sachs Wolfe effect. To our knowledge, this is the first time such an analysis has been done for an arbitrary $f(R)$ model. 

To study the evolution of various background and perturbed quantities, we have used the model-independent data of the late-time expansion history of the Universe constructed by Shafieloo et.al~\cite{2018-Shafieloo.etal-PRD}. We have analyzed in the redshift range $z=0$ to $1.2$. Due to the scarcity of PANTHEON data at redshift greater than $1.2$, we have not included the high redshift data in our analysis. We assumed that the effect of modifications to gravity begins to contribute from $z  = 1.2$ and have kept the success of standard general relativity at early times. The current analysis can be extended to higher redshifts once more data is available on the expansion history of the Universe at higher redshifts.
 
In Appendix (\ref{app:Background}), we have compared generic $f(R)$ model with the popular $f(R)$ models in the literature. We have shown that the evolution of $F(z)$ constructed can describe various $f(R)$ models to explain the late-time acceleration of the Universe.  
 
To keep the calculations transparent, we have assumed that pressureless matter contributes to the stress-tensor. Extending the analysis for multiple fields is possible. This is currently under investigation. 

Our analysis shows that the growth of $\Psi_{\rm MG}$ is less than $\Phi_{\rm MG}$ for $f(R)$ theories. It is interesting to see whether this feature is common for all modified gravity theories. This is currently under investigation. 

We have shown that the perturbation modes with larger wave number shows larger deviation of $\Psi_{MG}/ \Phi_{MG}$ from $\Psi_{GR} / \Phi_{GR} = 1$. For this parameter to be a tool for the detection of modified gravity theories, we need to confirm/infirm for other theories of gravity.

\section{Acknowledgements}
We thank Benjamin L'Huillier and Arman Shafieloo for providing the model-independent data on the evolution of the Hubble parameter and for their suggestions on using the data for this work. We would also like to thank Aseem Paranjape and Tarun Deep Saini for fruitful discussions. We are grateful to Navya Nagananda for her help during the initial stages of this work. We thank the anonymous referee for the insightful comments which helped the authors spot a bug in the numerical codes. JPS is supported by CSIR Senior Research Fellowship, India. The work is partially supported by IRCC Seed Grant, IIT Bombay.

\appendix
\section{Background Evolution in $f(R)$: Details} 
\label{app:Background}

In this Appendix, we provide more details of the results from Section (\ref{sec:Background}) for Scenario II. 

\subsection{Numerical Analysis in redshift space} 

The time derivatives in the evolution equations can be rewritten 
in terms of derivative with respect to redshift $z$, using the following relations
\begin{eqnarray}
\label{eq:dtdz1}
\dfrac{d}{dt} &=& -H (1 + z) \dfrac{d}{dz} \\
\label{eq:dtdz2}
\dfrac{d^2}{dt^2} &=& H (1 + z)^2 \dfrac{dH}{dz} \dfrac{d}{dz} + H^2 (1 + z)\dfrac{d}{dz} + H^2 (1 + z)^2 \dfrac{d^2}{dz^2}
\end{eqnarray}
To numerically solve the equations, derivatives are rewritten using the central difference method.
\begin{eqnarray}
\label{eq:centraldiff}
\dfrac{df(z)}{dz}&=& \dfrac{f_{i + 1} - f_{i - 1}}{2 dz}\\
\label{eq:centaldiff2}
\dfrac{d^2 f(z)}{dz^2} &=& \dfrac{f_{i + 1} - 2 f_i + f_{i -1}}{dz^2}
\end{eqnarray}

\subsection{Comparison of general $f(R)$ model with popular models}
\label{sec:fit}

Many $f(R)$ models have been proposed that lead to the late time acceleration of the Universe~\cite{2007-Hu.Sawicki-PRD,2007-Starobinsky-JETPLett,2007-Amendola.etal-PRD,2007-Appleby.Battye-PLB,2008-Tsujikawa-PRD}. 

In this Appendix, we corroborate the evolution of  $F$ we obtained using different realizations of the expansion history of the universe compared with two such $f(R)$ models. Table \ref{tab:fit} gives the best fit with root mean square error (RMS) for the constructed realizations of $F(z)$ in the range $0<z<1.2$ for these models. 

\begin{table}[!h]
\caption{\label{tab:example} Best fit for $f(R)$ models}
\begin{tabular}
{| p{1cm} | p{8cm}| p{8cm} | } \hline 
& ~~~~~~~~~~~~~~~~~~~~~~~~{\bf $F(R)$} & {\centering ~~~~~~~~~~~~~~~~~~~~~~~~~~\bf Best fit} \\ 
\hline
\centering 1 & 
$$
F(R) = 1 - 2\lambda n \dfrac{R}{R_0}\left[ 1+ \left(\dfrac{R}{R_0}\right)^2 \right]^{-(n+1)}
$$
\centerline{Starobinsky \cite{2007-Starobinsky-JETPLett}}
 &
\begin{equation*}
n=3.676, \quad
\lambda = 1.312 \times 10^6, \quad R_0 = H_0^2
\end{equation*}

\centerline{
$RMS = 6.8 \times 10^{-4}$}
 \\ 
\hline
\centering 2 & \begin{equation*}
\label{eq:Husawifit}
F(R) = 1 -  n \dfrac{c_1}{c2} \dfrac{\left(\dfrac{R}{R_0}\right)^{n-1}}{\left[\left(\dfrac{R}{R_0}\right)^n - 1 \right]^2}
\end{equation*}
\centerline{Hu \& Sawicki \cite{2007-Hu.Sawicki-PRD}}&

\begin{equation*}
n=7.176, \quad c_1/c_2=8.67*10^5, \quad R_0=H_0^2
\end{equation*}

\centerline{$
RMS = 6.6 \times 10^{-4}
$} \\ 
\hline
\end{tabular}
\label{tab:fit}
\end{table}
%

We see that the evolution of $F(z)$ constructed using the different realizations of the expansion history of the universe describes the $f(R)$ models that have been proposed in the literature.

\subsection{Evolution of $F(z)$ for different initial conditions}
\label{sec:Fpini}
As mentioned in Sec. (\ref{sec:Background}), we obtain the evolution of $\overline{F}(z)$ for different values of the initial condition. It is important to note that the physical assumption that at redshift $z = 1.2$ the gravity is described by 
GR leads to the condition that $\overline{F}(1.2) = 1$. Here, we show that the 
results obtained in Sec. (\ref{sec:Background}) do not depend on this value. In the plots below we have plotted the evolution of $\overline{F}$ for four other initial conditions:
\begin{equation}
 \overline{\frac{dF}{dz}} \Big|_{z=1.2} =  10^{-3}, 10^{-4}, 10^{-5}, 10^{-6}, 10^{-7}
\end{equation}
\begin{figure}[!h]
\includegraphics[scale=.45]{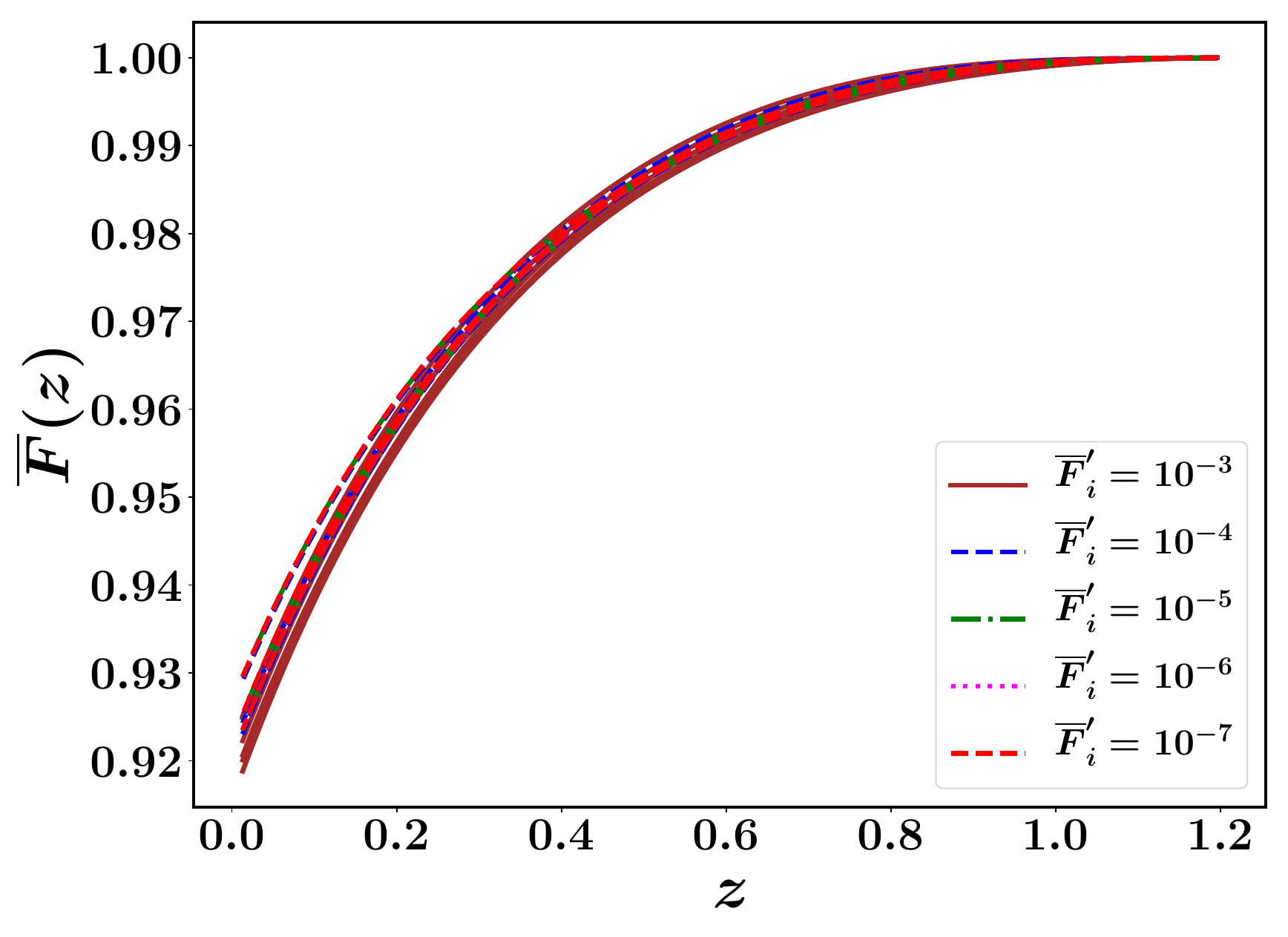}
\caption{Evolution of $\overline{F}$ for different values of $F'(z = 1.2)$}
\end{figure}

Here we see that the evolution of $\overline{F}$ does not vary depending on the initial conditions. One significant change is the number of datasets that satisfy the requirement $\overline{F}^{\prime}(z)>0$. More datasets satisfy this condition with larger value of $\overline{F}^{\prime}_i$. But for a given realization of the expansion history of the universe, the choice of initial conditions does not have any bearing on the evolution of $F(z)$. 

\begin{figure}[!h]
  \begin{minipage}[b]{0.5 \textwidth}
\includegraphics[scale=0.25]{F_0_3158_1_2.pdf} 
\end{minipage}\hfill
 \begin{minipage}[b]{0.5 \textwidth}
 \includegraphics[scale=0.25]{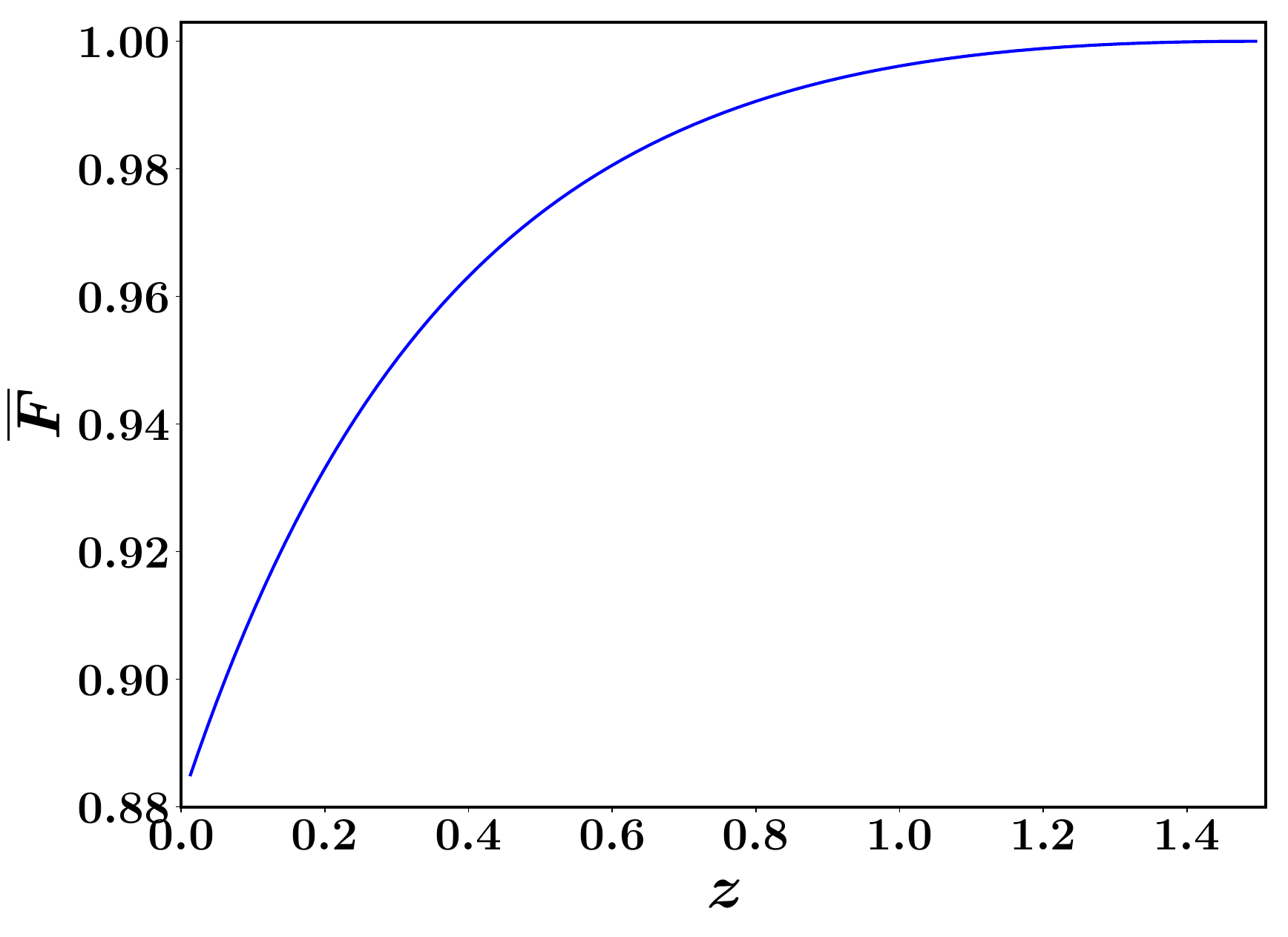}
 \end{minipage}
\caption{Plot of $\overline{F}~vs~z$. Left panel: $0 \leq z \leq 1.2$; Right panel: $0 \leq z \leq 1.5$.}
\end{figure}

\section{Evolution over a larger range of redshift}
In this appendix we compare the evolution of various background and perturbed quantities over the redshift ranges $z= 1.2 - 0$ and $z = 1.5 - 0 $. The number of datasets satisfying the condition $F'(z)>0$ changes with the redshift range over which these quantities are evolved.

 \begin{figure}[!h]
  \begin{minipage}[b]{0.5 \textwidth}
\includegraphics[scale=0.25]{phi_psi_0_3158_5_1_1_2.pdf} 
\end{minipage}\hfill
 \begin{minipage}[b]{0.5 \textwidth}
 \includegraphics[scale=0.25]{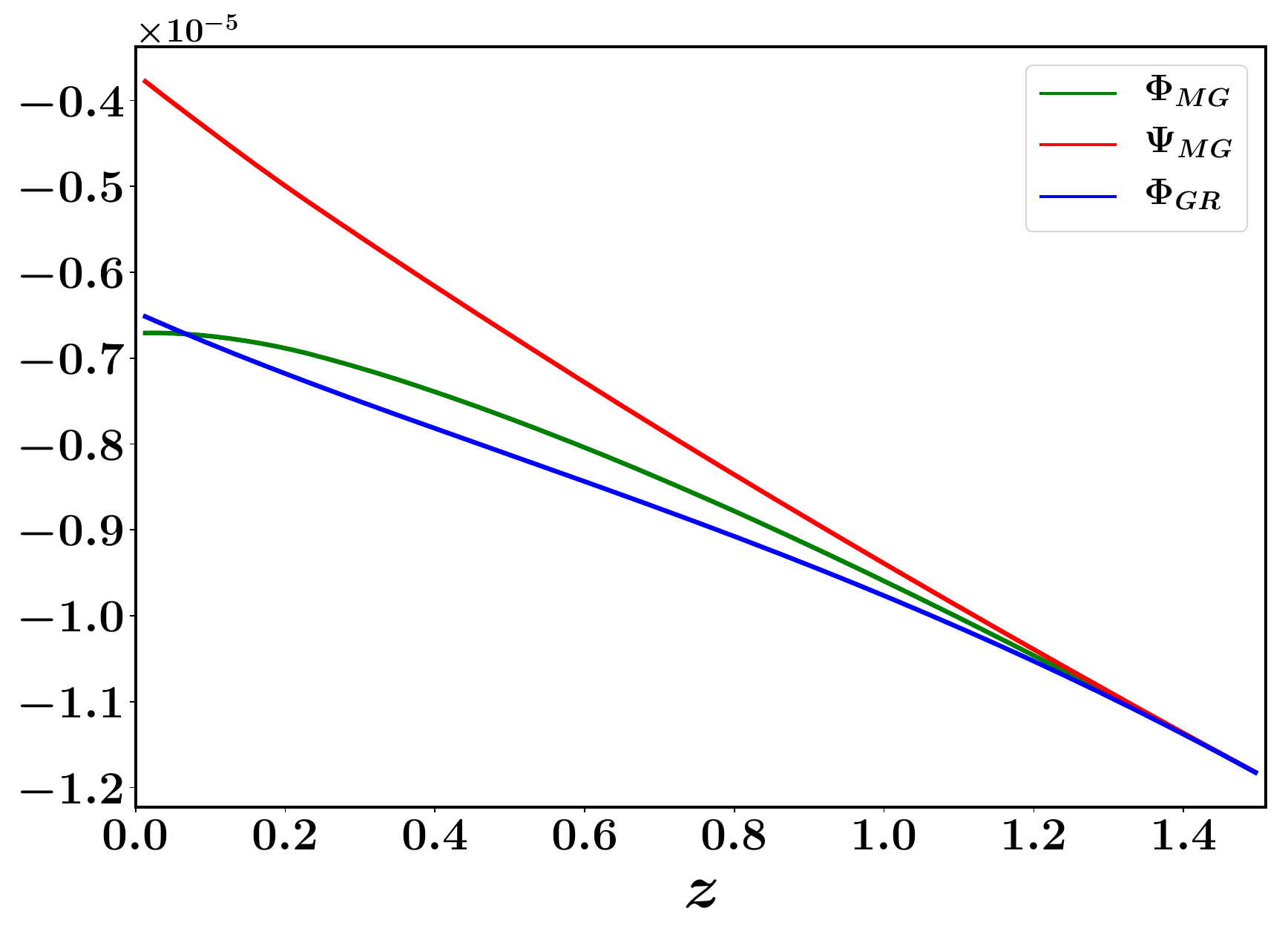}
 \end{minipage}
\caption{Plot of $\Phi_{MG},~\Psi_{MG},$ and $\Psi_{GR}~vs~z$. Left panel: $0 \leq z \leq 1.2$; Right panel: $0 \leq z \leq 1.5$.}
\label{fig:Psi_Phi_1_5}
\end{figure}
 \begin{figure}[!h]
  \begin{minipage}[b]{0.5 \textwidth}
\includegraphics[scale=0.25]{psi_by_phi_0_3158_5_1_1_2.pdf} 
\end{minipage}\hfill
 \begin{minipage}[b]{0.5 \textwidth}
 \includegraphics[scale=0.25]{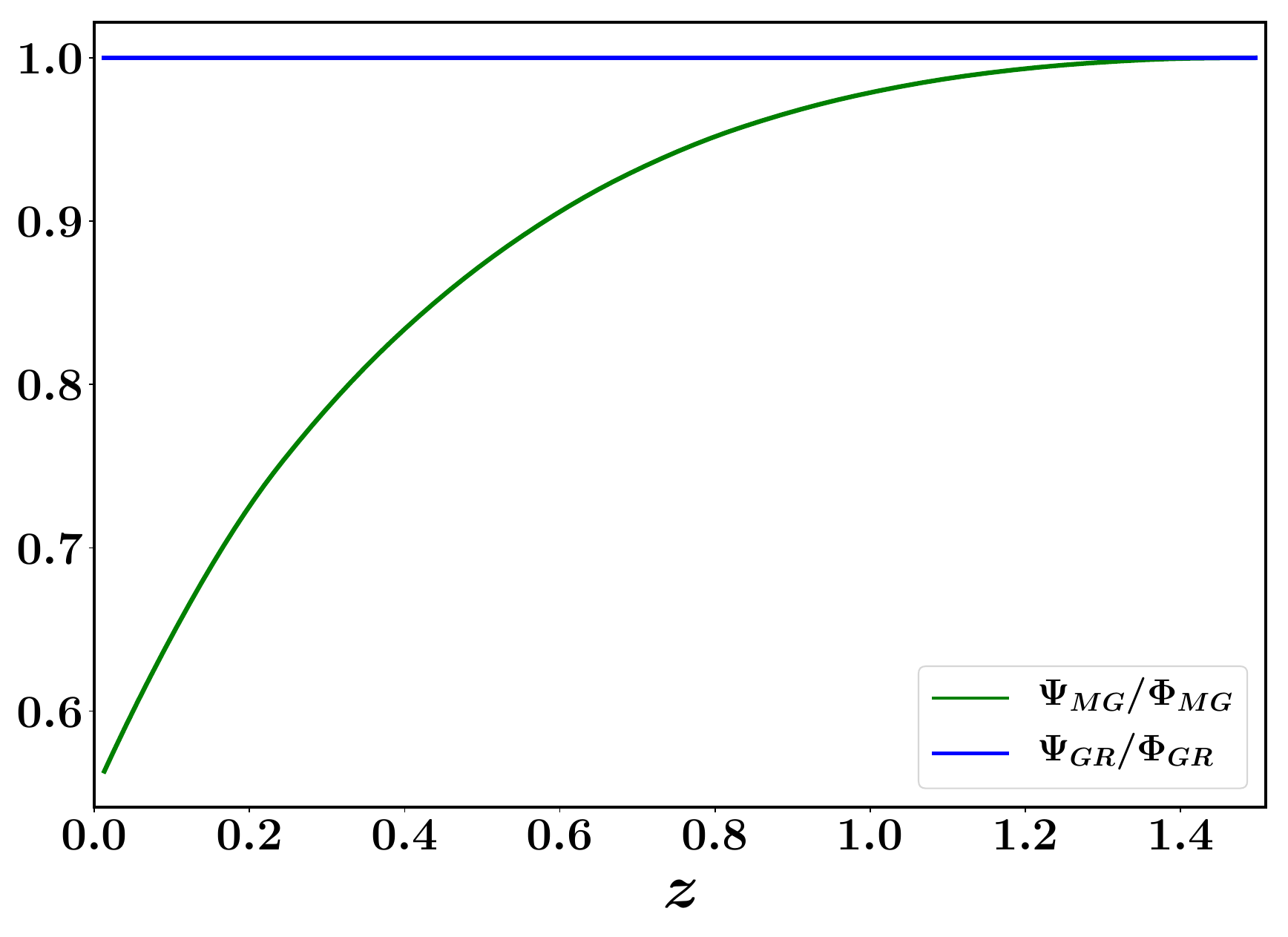}
 \end{minipage}
\caption{Plot of $\Psi_{MG}/\Phi_{MG},$ and $\Psi_{GR}/ \Psi_{GR}~vs~z$. Left panel: $0 \leq z \leq 1.2$; Right panel: $0 \leq z \leq 1.5$.}
\end{figure}

 \begin{figure}[!h]
  \begin{minipage}[b]{0.5 \textwidth}
\includegraphics[scale=0.25]{phi_plus_psi_0_3158_5_1_1_2.pdf} 
\end{minipage}\hfill
 \begin{minipage}[b]{0.5 \textwidth}
 \includegraphics[scale=0.25]{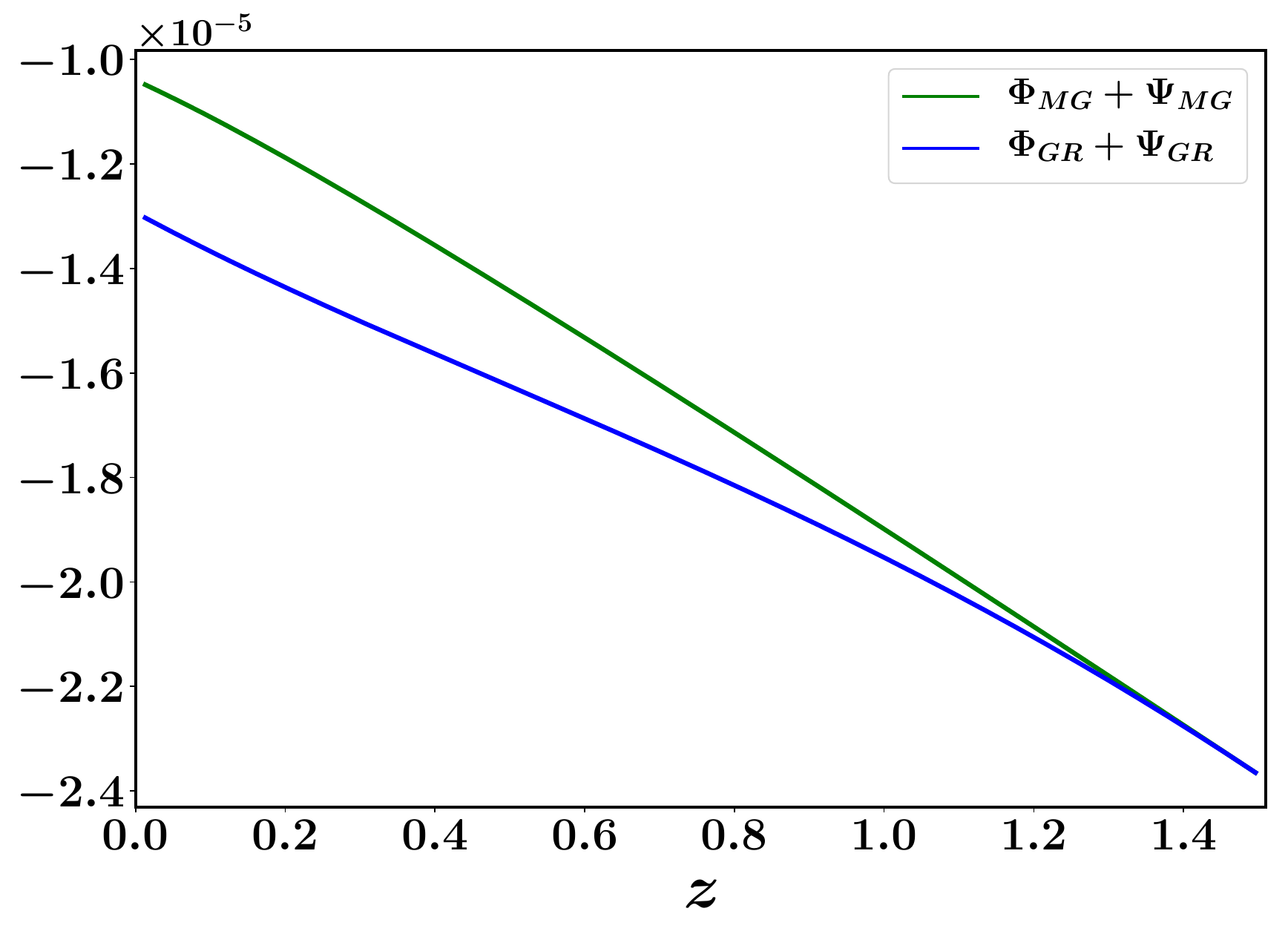}
 \end{minipage}
\caption{Plot of $\Psi_{MG}+\Phi_{MG},$ and $\Psi_{GR}+ \Psi_{GR}~vs~z$. Left panel: $0 \leq z \leq 1.2$; Right panel: $0 \leq z \leq 1.5$.}
\end{figure}

 \begin{figure}[!h]
  \begin{minipage}[b]{0.5 \textwidth}
\includegraphics[scale=0.25]{phi_plus_psi_prime_0_3158_5_1_1_2.pdf} 
\end{minipage}\hfill
 \begin{minipage}[b]{0.5 \textwidth}
 \includegraphics[scale=0.25]{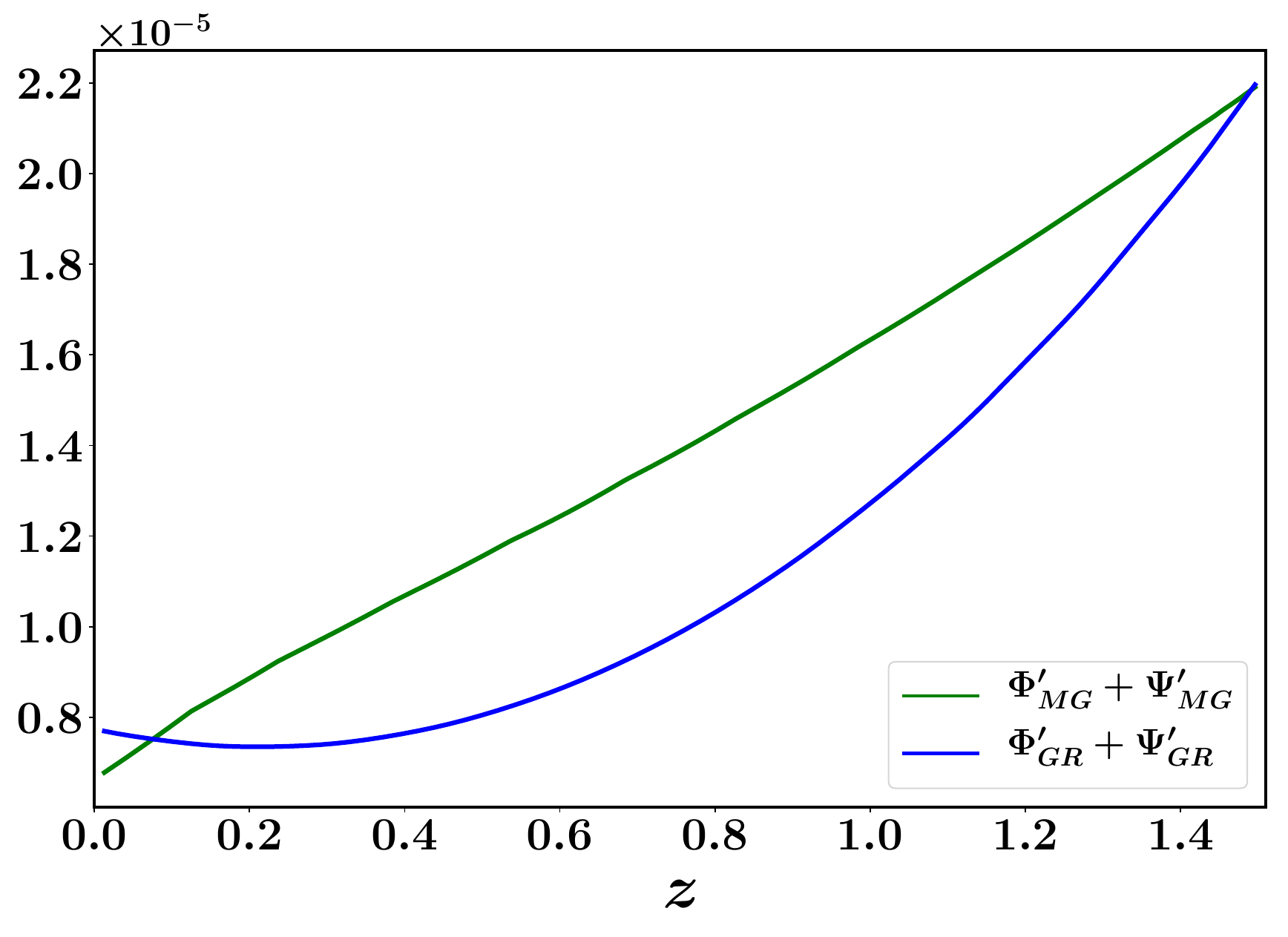}
 \end{minipage}
\caption{Plot of $\Psi_{MG}'+\Phi_{MG}',$ and $\Psi_{GR}'+ \Psi_{GR}'~vs~z$. Left panel: $0 \leq z \leq 1.2$; Right panel: $0 \leq z \leq 1.5$.}
\end{figure}

Here we see that as these quantities are evolved over a broader range of redshift, the difference between two scenarios become more significant. Hence, there will be an observable difference in the evolution of these quantities over a large redshift range.

\section{Simplified first order evolution equations for $\Phi$ and $\Psi$ in Scenario II}
\label{app:Scenario2}

In this appendix, we provide the complete set of first order scalar perturbation equations for an arbitrary $f(R)$. For the perturbed line-element (\ref{eq:FRWpert}), 
the modified Einstein's equations (\ref{eq:fRfieldeq}) and the trace equation (\ref{eq:Trace}) lead to:
%
\begin{eqnarray}
\label{eq:MGpert1}
\nonumber -\dfrac{\nabla^2 \Psi}{a^2} + 3H(H \Phi + \dot{\Psi}) +\dfrac{1}{2F}\biggl[ \left( 3H^2 + 3 \dot{H} + \dfrac{\nabla^2}{a^2} \right) \delta F -\\  3H \dot{\delta F} +  3H\dot{F} \Phi + 3 \dot{F} (H \Phi + \dot{\Psi}) + \kappa^2 \delta \rho  \biggr] &=& 0,
\end{eqnarray}
\begin{eqnarray}
\label{eq:MGpert3}
\nonumber 3 \bigl(\dot{H} \Phi + H \dot{\Phi} + \ddot{\Psi} \bigr) + 6H \bigl(H \Phi + \dot{\Psi} \bigr) + 3 \dot{H} \Phi + \dfrac{{\nabla}^2 \Phi}{a^2} - \dfrac{1}{2F} \biggl[ 3 \ddot{\delta F} + 3H \dot{\delta F} - 6H^2 
\\  \delta F  - \dfrac{{\nabla}^2 \delta F}{a^2} - 3 \dot{F} \dot{\Phi} - 3 \dot{F} \bigl(H \Phi + \dot{\Psi} \bigr) - \bigl(3 H \dot{F} + 6 \ddot{F} \bigr) \Phi + {\kappa}^2 \delta \rho \biggr]&=&0, 
\end{eqnarray}
\begin{eqnarray}
\label{eq:MGpert6} \nonumber
\ddot{\delta F} + 3 H \dot{\delta F} + \left(\frac{k^2}{a^2}-4 H^2 - 2 \dot{H}\right) \delta F
 -2 F \ddot{\Psi} -\left( 8 F H + 3 \dot{F} \right) \dot{\Psi}\\ -\left(2 F H + \dot{F} \right) \dot{\Phi}
 -\left( 6H \dot{F} + 2 \ddot{F} + 4 F \dot{H} +8 F H^2 - \frac{2Fk^2}{3 a^2}\right)\Phi
 -\frac{4Fk^2}{3a^2}\Psi - \frac{\kappa^2 \delta \rho}{3}&=& 0 \\
%
\label{eq:MGpert2}
H \Phi + \dot{\Psi} - \dfrac{1}{2F} \bigl(\dot{\delta F} - H \delta F - \dot{F} \Phi \bigr)&=&0, \\
\label{eq:MGpert4}
\Phi - \Psi  +\dfrac{ \delta F}{F}&=&0,\\
\label{eq:MGpert5}
\delta F - F^{\prime} \delta R&=&0,
\end{eqnarray}
Substituting for $\delta F$ using Eq.~(\ref{eq:MGpert4}), we get
\begin{eqnarray}
\label{eq:MGpert11} \nonumber
\left(3 H + \frac{3 \dot{F}}{F} \right) \dot{\Psi} + \left(3 H^2 + 3 \dot{H}
-3H \frac{\dot{F}}{F} + \frac{k^2}{a^2} \right) \Psi   \\ + \left( 3 H^2
 - 3 \dot{H} + 9 H \frac{\dot{F}}{F} + \frac{k^2}{a^2} \right) \Phi
 + 3 H \dot{\Phi} + \frac{\kappa^2 \delta \rho}{F} &=& 0,\\
\label{eq:MGpert21}
\dot{\Phi} + \dot{\Psi} + \left(H -\frac{\dot{F}}{F}\right) \Psi +\left(H+ 2 \frac{\dot{F}}{F}\right)\Phi &=& 0, \\
\label{eq:MGpert31}
\nonumber\ddot{\Psi} +\ddot{\Phi}+ 3\left(H+  \frac{\dot{F}}{F}\right)\dot{\Phi}+ \left(3H - \frac{\dot{F}}{F}\right)\dot{\Psi}+
 \left( 2 H^2 - \frac{\ddot{F}}{F}- \frac{H \dot{F}}{F} - \frac{k^2}{3 a^2} \right) \Psi
 \\ +\left( 2 H^2 + \frac{3\ddot{F}}{F}+ \frac{3H \dot{F}}{F} +4 \dot{H}- \frac{k^2}{3 a^2}\right)\Phi 
 -\frac{\kappa^2 \delta \rho}{3 F}&=&0,
\end{eqnarray}
\begin{eqnarray}
\label{eq:MGpert61}
\nonumber \ddot{\Phi}+\ddot{\Psi} + \left(5 H +\frac{\dot{F}}{F} \right)\dot{\Psi} +
\left( 4 H^2 - \frac{\ddot{F}}{F}- \frac{3H \dot{F}}{F} +2 \dot{H}+ \frac{k^2}{3 a^2}\right)\Psi \\+  \left(5 H +\frac{3\dot{F}}{F} \right)\dot{\Phi}
+ \left( 4 H^2 + \frac{3\ddot{F}}{F}+ \frac{9H \dot{F}}{F} +2 \dot{H}+ \frac{k^2}{3 a^2}\right) \Phi+
\frac{\kappa^2 \delta \rho}{3 F} &=&0
\end{eqnarray}
Substituting $\dot{\Phi}$ in Eq.~(\ref{eq:MGpert21}) using Eq.~(\ref{eq:MGpert11}), we get
\begin{equation}
 \label{eq:Psi}
 \dot{\Psi}+\left(H-\frac{F \dot{H}}{\dot{F}}+\frac{F}{3 \dot{F}}\frac{k^2}{a^2} \right)\Phi+
 \left(\frac{F \dot{H}}{\dot{F}}+\frac{F}{3 \dot{F}}\frac{k^2}{a^2}\right)\Psi 
 +\frac{\kappa^2 \delta \rho}{3 \dot{F}}=0 \, .
\end{equation}
Similarly, substituting $\ddot{\Phi}$ and $\dot{\Psi}$ in 
Eq.~(\ref{eq:MGpert61}) using Eq.~(\ref{eq:MGpert31}) and Eq.~(\ref{eq:Psi}), respectively, we get:
\begin{equation}
 \label{eq:Phi}
 \dot{\Phi}+\left(H-\frac{\dot{F}}{F}-\frac{F \dot{H}}{\dot{F}}-\frac{F}{3 \dot{F}}\frac{k^2}{a^2} \right)\Psi+
 \left(2 \frac{\dot{F}}{F}+\frac{F \dot{H}}{\dot{F}}-\frac{F}{3 \dot{F}}\frac{k^2}{a^2}\right)\Phi 
 -\frac{\kappa^2 \delta \rho}{3 \dot{F}}=0 \, .
\end{equation}

%

%

\end{document}